\documentclass[twocolumn, twocolappendix, astrosymb]{aastex701}

\graphicspath{{./}{figures/}}

\usepackage{graphicx}	% Including figure files
\usepackage{amsmath}	% Advanced maths commands
\usepackage{xspace}
\usepackage{tabularx}
\usepackage{booktabs}

\newcommand{\hi}{${\rm H}$\,{\sc i}\xspace}

\newcommand{\gr}[0]{\ensuremath{g\! - \!r}\xspace}
\newcommand{\mpch}[0]{\textit{h}$^{-1}$ cMpc\xspace}
\newcommand{\hmpc}[0]{\textit{h} cMpc$^{-1}$\xspace}
\newcommand{\hib}{\hi $\times$ Blue\xspace}
\newcommand{\hir}{\hi $\times$ Red\xspace}
\newcommand{\hia}{\hi $\times$ Galaxy\xspace}
\newcommand{\shi}{\ensuremath{_{\mathrm{HI}}\xspace}}
\newcommand{\xmA}{\textit{mK+cB}\xspace}
\newcommand{\xmB}{\textit{tK+cB}\xspace}
\newcommand{\xmC}{\textit{tK+cB+FoG}\xspace}
\newcommand{\xmD}{\textit{tK+sB+FoG}\xspace}
\newcommand{\amA}{\textit{cB}\xspace}
\newcommand{\amB}{\textit{sB}\xspace}
\newcommand{\amC}{\textit{cB+FoG}\xspace}
\newcommand{\amD}{\textit{sB+FoG}\xspace}
\newcommand{\kten}{\ensuremath{k_{10\%}\xspace}}
\newcommand{\dfog}[1]{\ensuremath{D_{\mathrm{FoG #1}}}}
\newcommand{\keff}{\ensuremath{k_{\mathrm{eff}}}\xspace}
\newcommand{\obr}{\ensuremath{\Omega\shi b\shi r_{\mathrm{HI-Gal}}}\xspace}
\newcommand{\hydro}{\textit{Hydro}\xspace}
\newcommand{\nbody}{\textit{N-body}\xspace}
\newcommand{\anal}{\textit{Analytical}\xspace}

\defcitealias{villaescusa-navarro_ingredients_2018}{VN18}
\defcitealias{wolz_h_2022}{W22}
\defcitealias{cunnington_h_2022}{C22}
\defcitealias{2023_chime_citation}{CHIME23}
\defcitealias{wang_breakdown_2021}{W21}
\defcitealias{spinelli_atomic_2020}{S20}
\defcitealias{paul_first_2023_eprint}{P23}
\defcitealias{Carucci24}{C24}

\begin{document}

\title{Lost in the FoG: Pitfalls of Models for Large-Scale Hydrogen Distributions}

\author[0000-0003-4414-5589]{Calvin K. Osinga}
\affiliation{Department of Astronomy, University of Maryland - College Park, 4296 Stadium Dr, 20742, College Park, MD, USA}
\email{calvinosinga@gmail.com}

\author{Benedikt Diemer}
\affiliation{Department of Astronomy, University of Maryland - College Park, 4296 Stadium Dr, 20742, College Park, MD, USA}
\email{diemer@umd.edu}

\author{Francisco Villaescusa-Navarro}
\affiliation{Center for Computational Astrophysics, Flatiron Institute, 162 5th Avenue, 10010, New York, NY, USA}
\email{fvillaescusa@princeton.edu}

\begin{abstract}

Large-scale \hi surveys and their cross-correlations with galaxy distributions have immense potential as cosmological probes. Interpreting these measurements requires theoretical models that must incorporate redshift-space distortions (RSDs), such as the Kaiser and fingers-of-God (FoG) effect, and differences in the tracer and matter distributions via the tracer bias. These effects are commonly approximated with assumptions that should be tested on simulated distributions. In this work, we use the hydrodynamical simulation suite IllustrisTNG to assess the performance of models of $z \leq 1$ \hi auto and \hi-galaxy cross-power spectra, finding that the models employed by recent observations introduce errors comparable to or exceeding their measurement uncertainties. In particular, neglecting FoG causes $\gtrsim 10\%$ deviations between the modeled and simulated power spectra at $k \gtrsim 0.1$ \hmpc, larger than assuming a constant bias which reaches the same error threshold at slightly smaller scales. However, even without these assumptions, models can still err by $\sim 10\%$ on observationally relevant scales. These remaining errors arise from multiple RSD damping sources on \hi clustering, which are not sufficiently described with a single FoG term. Overall, our results highlight the need for an improved understanding of RSDs to harness the capabilities of future measurements of \hi distributions.

\end{abstract}


\section{Introduction} \label{sec:intro}

% Start with overview of the role of models. I liked starting this way more since it gives the reader a better idea of the scope of the work immediately rather than the previous version of the introduction.
Cosmology imprints on the large-scale structure of the Universe, allowing measurements of the distribution of matter to probe the nature of dark matter and dark energy and constrain cosmological parameters \citep{jenkins_evolution_1998, eisenstein_detection_2005, reddick_cosmological_2014, DESI_IV_2025}. However, we cannot observe the Universe's large-scale structure directly. We instead rely on surveys of baryonic tracers such as atomic neutral hydrogen (\hi) or galaxies to infer the underlying matter distribution \citep{DESI_III_2024}, a process which requires models of the matter-tracer relationship. These models are simple in the linear regime where tracers faithfully follow the matter distribution, but baryonic physics and nonlinear effects complicate the behavior on small scales, demanding more sophisticated models. Despite the challenge, the stronger constraining power on small scales makes the development of these complex models worthwhile \citep{krause_cosmolike_2017, chisari_core_2019}.

% an assortment of models of the distribution of tracers. Notable examples, explanations. Phenomenological models favored due to relative simplicity.
Consequently, a large body of work has been dedicated to improving our understanding of structure in the quasi-linear and nonlinear regime. Notable examples of modeling techniques include Lagrangian perturbation theory \citep[LPT,][]{bernardeau_large-scale_2002, carlson_convolution_2013, vlah_lagrangian_2015} and halo occupation distribution (HOD) models \citep{zehavi_departures_2004, zheng_galaxy_2007}. However, both techniques have limitations: LPT does not natively include baryonic effects and HOD requires knowledge and assumptions of how tracers occupy halos. In certain applications, a simpler and more general prescription which assumes parameterized forms for the various nonlinear contributions is preferred.

% For example, recent observations of the HI-galaxy cross-correlation used phenomenological models to obtain constraints of the cosmic abundance of hydrogen. Why HI and galaxies are common tracers, why they are cross-correlated.
These general prescriptions have interpreted measurements of \hi and galactic distributions to constrain cosmological parameters. Of the two tracers, galaxies are the traditional probe of large-scale structure \citep[e.g.,][]{cole_2df_2005}, but \hi has arisen as a promising candidate due to its ability to cheaply and quickly observe large volumes of space via 21cm intensity mapping \citep{bharadwaj_using_2001, battye_neutral_2004, leo_constraining_2020}. However, 21cm intensity mapping experiments still struggle with low signal-to-noise ratios, arising from galactic foregrounds and systematics \citep{matteo_radio_2002}, making the interpretation of \hi auto power spectra challenging \citep[][hereafter \citetalias{paul_first_2023_eprint}]{paul_first_2023_eprint}. These noise sources are absent in galaxy surveys, so the noise can be reduced by cross-correlating \hi and galaxies. These cross-correlations have constrained the cosmological abundance of \hi ($\Omega\shi$) at low redshifts \citep[$z \leq 1$,][]{chang_intensity_2010, masui_measurement_2013, switzer_determination_2013, anderson_low-amplitude_2018}. In particular, \citet[][hereafter \citetalias{wolz_h_2022}]{wolz_h_2022}, \citet[][\citetalias{cunnington_h_2022}]{cunnington_h_2022}, \citet[][\citetalias{2023_chime_citation}]{2023_chime_citation}, and \citet[][\citetalias{Carucci24}]{Carucci24} have produced $\Omega\shi$ constraints competitive to other \hi observation techniques. As measurement and data processing techniques improve \citep{cunnington_impact_2019, carucci_recovery_2020, podczerwinski_needlet_2024}, \hi-galaxy cross-correlations will probe other cosmological quantities in the near future \citep{MeerKLASS_2024_eprint}.

However, the resulting constraints depend on how well the adopted models capture the relationship between observed tracer and matter distributions. We can evaluate these models' performance by comparing their predictions to known \hi, galaxy, and matter distributions in simulations. Cosmological hydrodynamical simulations now produce realistic low-redshift \hi and galaxy distributions in sufficiently large volumes to be useful for this purpose \citep{bahe_distribution_2016,nelson_first_2018, diemer_atomic_2019, stevens_atomic_2019, dave_galaxy_2020}. In particular, \citet{osinga_atomic_2024} showed that the \hi auto- and cross-correlations from IllustrisTNG agree closely with relevant observations at $z \leq 1$. We leverage the agreement between IllustrisTNG and observations to gain insight into the performance of large-scale \hi models in two phases. First, we report IllustrisTNG's predictions for the various model components and assess how well their simplified forms in the general model agree with the actual values. Second, we evaluate the accuracy of popular large-scale prescriptions by comparing their predictions to the correlations computed directly from IllustrisTNG.

The paper is structured as follows. We first describe the data from IllustrisTNG in Section~\ref{sec:methods}. Next, we explore general model designs in Section~\ref{sec:models}. Then, we describe predictions for each model ingredient individually in Section~\ref{sec:ing} and the models as a whole in Section~\ref{sec:perf}. We discuss directions for future work in Section~\ref{sec:disc}, before concluding in Section~\ref{sec:conc}. For brevity, some figures are referenced but not included; these are provided in the author's website\footnote{\url{www.calvinosinga.com/papers/hi_cosmo/sup_analysis}}.

\section{Simulation data} \label{sec:methods}

We use the IllustrisTNG suite of cosmological magneto-hydrodynamics simulations \citep{nelson_first_2018, pillepich_first_2018, springel_first_2018, naiman_first_2018, marinacci_first_2018, nelson_illustristng_2019}. This suite offers simulations in three different box sizes, 35 \mpch, 75 \mpch, and 205 \mpch, at varying resolutions, using the \citet{planck_collaboration_proper_citation} cosmological parameters ($\Omega_{\rm{m}} = 0.3089$, $\Omega_{\rm{b}} = 0.0486$, $h = 0.6774$, $\sigma_{\rm{s}} = 0.8159$).

The simulation is evolved using the AREPO code \citep{springel_e_2010}, which employs a tree-PM method for gravity calculations and a Voronoi mesh-based Godunov scheme for magneto-hydrodynamics. The IllustrisTNG simulations incorporate sub-grid models to simulate unresolved processes like star formation, gas cooling, and AGN \citep{vogelsberger_model_2013, weinberger_supermassive_2018}. The models are calibrated against a selection of observational data to accurately represent the galaxy population at low redshifts \citep{pillepich_simulating_2018}. Dark matter haloes are identified using the Friends-of-Friends (FoF) algorithm \citep{davis_evolution_1985}, and their internal substructures, or ``subhalos'', are cataloged using the \textsc{Subfind} algorithm \citep{springel_populating_2001}.

Our main analyses focus on the largest box, 205 \mpch (TNG300), at redshifts $z = 0, 0.5,$ and $z = 1$. We discuss the impact of resolution in Section~\ref{sec:disc} and directly compare our results to the 75 \mpch box (TNG100) in Appendix \ref{app:resolution}. To limit the impact of poorly resolved galaxies, we restrict our galaxy population to those containing at least 200 stellar particles, corresponding to a minimum stellar mass of $\approx 10^9 M_\odot$. We limit our analysis to redshifts $z = 0, 0.5,$ and $1$ where the color distribution in IllustrisTNG matches observations. The agreement worsens at earlier redshifts because IllustrisTNG lacks red, star-forming, dusty galaxies \citep{donnari_star_2019, gebek_atomic--molecular_2023, Gebek25}. 

We separate our galaxy sample into blue and red populations to mimic how galaxies are selected in optical surveys, as they detect galaxy samples visible in certain optical bands such as blue emission-line galaxies (ELGs) and luminous red galaxies (LRGs) \citep{alam_completed_2021}. We approximate these populations by categorizing galaxies as red or blue with their rest-frame \gr color values \citep{stoughton_sloan_2002}: galaxies with $\gr \geq 0.6$ are blue, with all others being red. This threshold changes with time to reflect the overall color evolution of the galaxy population. Specifically, \gr thresholds are set at 0.6, 0.55, and 0.5 for redshifts 0, 0.5, and 1, respectively. While testing various color definitions with and without dust corrections \citep{nelson_first_2018}, \citet{osinga_atomic_2024} found that these adjustments only substantially impact small-scale clustering at $k \gtrsim 1$ \hmpc (see their appendix A). A more exact method of matching IllustrisTNG galaxies to ELGs and LRGs will only somewhat impact the quantitative results from Section~\ref{sec:ing} (e.g., Tables~\ref{tab:bias}-\ref{tab:sigma}). The scale-dependent behavior in Section~\ref{sec:ing} and overall model performance presented in Section~\ref{sec:perf} are insensitive to small changes in the definitions of blue and red galaxies.

The final tracer, \hi, is not explicitly modeled in IllustrisTNG, so we must turn to post-processing to separate atomic and molecular hydrogen \citep{diemer_modeling_2018}. These post-processed distributions accurately reproduce a number of observed \hi relationships \citep{stevens_atomic_2019}. For example, simulated \hi rotation velocities agree with observations on average \citep{goddy_comparison_2023} despite some increased asymmetry \citep{watts_global_2020}. \hi surface densities also match, although \hi-rich galaxies in IllustrisTNG appear to have suppressed \hi abundance in their centers \citep{diemer_atomic_2019}. We expect these small differences to be negligible in our results. We have tested a wide range of post-processing models, but find negligible difference between the different treatments \citep[an offset of $\lesssim 2\%$ in amplitude, see][]{osinga_atomic_2024}, consequently we restrict our analysis to the single model from \citet{villaescusa-navarro_ingredients_2018}, hereafter \citetalias{villaescusa-navarro_ingredients_2018}.

\section{Models of large-scale \hi distributions} \label{sec:models}

Large-scale structure models are designed to infer the underlying matter distribution from the auto or cross-correlations of observed tracers. To accomplish this task, these models must properly handle many effects that can generally be categorized into three groupings: nonlinearities, tracer behavior, and redshift-space distortions (RSDs). The first grouping stems from gravitational instabilities complicating analytical descriptions of matter distributions. The second arises from the need to ascertain the entire matter distribution from a small subset of visible matter. The final grouping, RSDs, emerges from line-of-sight velocities displacing the apparent positions of observed tracers. In Sections~\ref{model:nonlin}-\ref{models:rsd}, we describe the modeling of nonlinearities, tracer behavior, and RSDs before presenting the general form for the models we test in this work in Section~\ref{models:complete}. We conclude in Section~\ref{models:assump} by highlighting the model assumptions relevant to our analysis in Sections~\ref{sec:ing}-\ref{sec:perf}. 

\begin{figure}
    \centering
    \includegraphics[width=.85\linewidth]{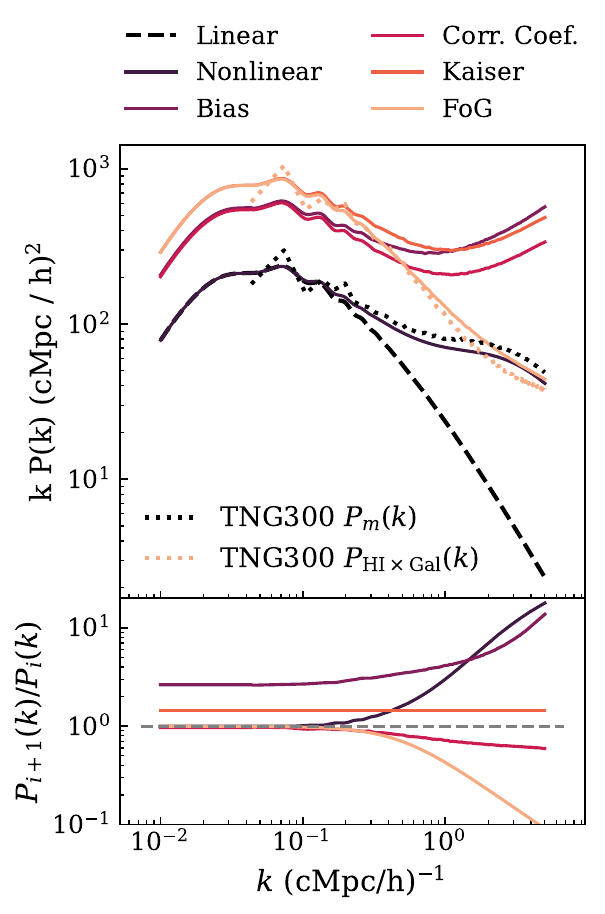}
    \caption{A deconstructed example \hi-galaxy cross-power spectrum, providing insight into the contribution of each model ingredient. The linear matter power spectrum is shown with the black dashed line in the top panel, with the wiggles at $k \sim 0.05$ \hmpc corresponding to baryonic acoustic oscillations. We then add each ingredient from Equation~\ref{eq:cross_full} to the linear power spectrum sequentially, lightening the color as each is added (see text for ingredient descriptions). The $z = 1$ matter power spectrum (black) and \hi-galaxy cross-power spectrum (salmon) for TNG300 are shown in dotted lines, in fair agreement with the corresponding model values. The bottom panel shows the ratio of the power spectra with and without that component in order to more easily visualize its individual contribution.}
    \label{fig:demo}
\end{figure}

Throughout this section, we introduce model ingredients and isolate their contributions to a key summary statistic that characterizes 3D distributions called the power spectrum, presented in Fig.~\ref{fig:demo}. The power spectrum describes how the variance of a field like the density contrast ($\delta (\boldsymbol{x}) = \rho(\boldsymbol{x}) / \overline{\rho} - 1$) changes across different spatial scales. Mathematically, the power spectrum is defined as
\begin{equation}
\label{eq:pk}
    \langle \Tilde{\delta}_i(\boldsymbol{k}) \Tilde{\delta}_j(\boldsymbol{k}') \rangle = (2\pi)^3 P_{i \times j}(k) \delta_D^3(\boldsymbol{k} + \boldsymbol{k}') \,.
\end{equation}
Here, $\delta_D$ is the Dirac delta function and $\Tilde{\delta_i}$ represents the Fourier transform of the overdensity. $P_{i \times j}(k)$ is the power spectrum and $\boldsymbol{k}$ is the wavenumber, with bold denoting a vector. $P_{i \times j} (\boldsymbol{k}) = P_{i \times j} (k)$ due to the cosmological principle. If $i=j$ in the above equation, $P_{i \times j}(k)$ is called an auto power spectrum, denoted $P_i(k)$. Otherwise, $P_{i \times j}(k)$ is a cross-power spectrum, capturing the correlation between different fields or populations. $i$ and $j$ represent general populations, but in this work we strictly consider \hi-galaxy cross-power spectra.

\subsection{Nonlinearities} \label{model:nonlin}

At large scales, prescriptions are relatively simple since the growth of the small large-scale perturbations ($\delta \ll 1$) can be described linearly \citep[for derivation, see][]{dodelson_modern_2020}. However, this growth becomes increasingly nonlinear toward small scales until gravitational instabilities form collapsed structures. On these scales, we rely on numerical methods to predict the matter power spectrum.

The transition between the linear and nonlinear regime can be found by comparing their power spectra in Fig.~\ref{fig:demo}. We use \textsc{camb}  \citep{lewis_efficient_2000, lewis_camb_2011} and \textsc{halofit} \citep{takahashi_revising_2012, smith_precision_2019} to determine the linear ($P_{\mathrm{L}}$) and nonlinear ($P_{\mathrm{NL}}$) power spectra, respectively. The two diverge at $k \sim 0.1$ \hmpc at $z = 1$, which agrees with other works studying the linear-nonlinear transition \citep[e.g.,][]{smith_stable_2003}. Roughly at the same scales, the wiggles from baryonic acoustic oscillations \citep{eisenstein_detection_2005} imprint on both the linear and nonlinear power spectra.

We also compare the nonlinear power spectra calculated from TNG300 and \textsc{halofit}. At large scales, the two roughly agree, although the small number of modes at the largest scales that TNG300 probes causes statistical variations that fall above and below the \textsc{halofit} result (we address this in Section~\ref{ing:bias}). At smaller scales, \textsc{halofit} and TNG300 diverge, with TNG300 having slightly more power, although the shapes match closely. Differences are expected since the \textsc{halofit} results are tuned to dark matter-only N-body simulations and the TNG300 power spectra include baryonic effects and are subject to cosmic variance \citep[][]{miller_detecting_2002, somerville_cosmic_2004, van_daalen_effects_2011, li_super-sample_2014, springel_first_2018}. For demonstrative purposes, we will use the \textsc{halofit} matter power spectrum for this section but strictly use the TNG300 result for the remainder of the paper. 

\subsection{From matter to tracers} \label{models:tracer} 

The matter power spectrum described in the previous section is not directly observable and must be inferred from measurements of visible tracers. The quantity that maps from tracer to matter overdensities is called the bias, yielding the following relationship:
\begin{equation}
\label{eq:bias}
    P_i (k) = b_i^2 (k) P_{\rm{m}} (k) + P_{\rm{SN}}\,.
\end{equation}
Here, $i$ represents some tracer population, which will always be \hi or a subsample of galaxies in our case. We incorporate the shot noise ($P_{\rm{SN}}$) into our bias definition, which is a common choice in observations --- either version does not impact the overall model performance but can change to what terms errors are attributed (see Appendix~\ref{app:SN}).

The impact of the \hi and galaxy biases is shown with the purple line in Fig.~\ref{fig:demo}, which corresponds mathematically to $b\shi (k) b_{\mathrm{Gal}}(k) P_{\mathrm{NL}} (k)$. The ratios in the bottom panel of Fig.~\ref{fig:demo} show that both biases are constant on large scales until $k \sim 0.1$ \hmpc where they increase. The galaxy and \hi biases inherit this shape from the halo bias --- halos tend to occupy the most nonlinear regions of space and thus are increasingly more clustered than matter on small scales, such that the halo bias increases \citep{jeong_perturbation_2009, nishimichi_scale_2012, nishizawa_perturbation_2013, paranjape_bias_2013}. Nearly all galaxies and \hi inhabit halos at $z \leq 1$ \citep[][\citetalias{villaescusa-navarro_ingredients_2018}]{white_core_1978} such that their biases mirror the scale-dependence of the halo bias. The \hi and galaxy biases used in Fig.~\ref{fig:demo} are estimated from TNG300 (discussed more in Section~\ref{sec:ing}) --- for scales beyond what TNG300 probes, we assume that the biases are constant and extrapolate the large-scale values from TNG300.

The bias sufficiently describes the matter-tracer relationship in auto power spectra, but cross-power spectra require an additional term: the correlation coefficient. The correlation coefficient measures the stochasticity in the relationship between tracers of the cosmic density field, with one, zero, and negative one corresponding to completely correlated, random, and anti-correlated distributions, respectively. Mathematically, the correlation coefficient is defined as
\begin{equation} \label{eq:corr_coef}
    r_{i-j} (k) = \frac{P_{i \times j} (k)}{\sqrt{P_i (k) P_j (k)}} \,.
\end{equation}
Similarly to the bias, the correlation coefficient is expected to be constant on large scales and approach zero toward small scales as stochasticity becomes more significant \citep{carucci_cross-correlation_2017, osinga_atomic_2024}. These trends are shown with the maroon line in Fig.~\ref{fig:demo}, which represents the expression $b\shi (k) b_{\mathrm{Gal}}(k)  r_{\mathrm{HI - Gal}} (k) P_{\mathrm{NL}} (k)$ and completely describes the tracer-tracer and tracer-matter relationships. We use the \hi-galaxy correlation coefficient from TNG300, assuming that $r_{\mathrm{HI - Gal}} = 1$ at scales larger than what TNG300 probes. The correlation coefficient suppresses the power spectrum at $k \gtrsim 0.2$ \hmpc, with this effect strengthening toward small scales. Overall, the contribution of the correlation coefficient is fairly small relative to the other components, reducing the power spectrum by a factor of 2 at the smallest scales shown in Fig.~\ref{fig:demo}.

\subsection{Incorporating redshift-space distortions} \label{models:rsd}

RSDs arise from velocities along the line of sight shifting the apparent positions of \hi and galaxies, such that the mapping between a tracer's real- ($\boldsymbol{r}$) and redshift-space ($\boldsymbol{s}$) positions is
\begin{equation} \label{eq:sr_mapping}
    \boldsymbol{s} = \boldsymbol{r} + \frac{v_z(\boldsymbol{r})}{a H(z)} \Hat{\boldsymbol{z}} \,.
\end{equation} 
Here, $\Hat{\boldsymbol{z}}$ is the line of sight and $v_z$ represents the line-of-sight velocities. Density conservation\footnote{Strictly speaking, intensity mapping considers the specific intensity, not density, but the derivation follows a similar procedure regardless \citep{bharadwaj_using_2001, bharadwaj_cosmic_2004}.} implies that the real- and redshift-space density fields are related with
\begin{equation} \label{eq:dsr_mapping}
    \delta^S = \left\lvert \frac{\partial \boldsymbol{s}}{\partial \boldsymbol{r}}\right\rvert^{-1} (1 + \delta^R) - 1 \,.
\end{equation}
The $R$ and $S$ superscripts denote quantities in real and redshift space, respectively. Taking the Fourier transform of Equation~\ref{eq:dsr_mapping} and expressing the mapping in terms of the power spectra yields
\begin{multline} \label{eq:rsd_exact}
    P^S (k) = \int d^3x e^{i \boldsymbol{k} \cdot \boldsymbol{x}} \langle e^{-ik\mu f \Delta u_z} \times \\(\delta (\boldsymbol{r}) + f \nabla_z u_z (\boldsymbol{r}))(\delta (\boldsymbol{r}') + f \nabla_z u_z(\boldsymbol{r}')\rangle  \,.  
\end{multline}
Here, $f = d \ln D(a) / d \ln a$, with $D(a)$ defined as the linear growth rate \citep{dodelson_modern_2020}. $u_z$ represents the renormalized line-of-sight velocities $u_z(\boldsymbol{r}) = -v_z (\boldsymbol{r}) / (a H f)$. $\boldsymbol{x} = \boldsymbol{r} - \boldsymbol{r}'$ is the separation between two points. $\langle ... \rangle$ represents the ensemble average. The remaining term is the difference between local line-of-sight velocities $\Delta u_z = u_z (\boldsymbol{r}) - u_z (\boldsymbol{r}')$. For more information on Equation~\ref{eq:rsd_exact}, see \citet{taruya_baryon_2010}. 

While Equation~\ref{eq:rsd_exact} is challenging to evaluate exactly, it can be simplified at linear scales. In this regime, the density and velocity fields are directly related via continuity ($\delta_L + \nabla \cdot \boldsymbol{v}_L = 0$), leading to the famous \citet{kaiser_clustering_1987} result:
\begin{equation} \label{eq:kaiser_matter}
    P^S_{\mathrm{m}} (k, \mu) = (1 + f \mu^2)^2 P^R_{\mathrm{m}} (k) \,.
\end{equation}
In this equation, $\mu = \Hat{\boldsymbol{k}} \cdot \Hat{\boldsymbol{z}}$, which is a common way to parameterize line-of-sight anisotropies. We use the subscript $m$ to denote quantities for the distribution of matter. The Kaiser effect describes the increase in clustering from the coherent movement of large-scale structures, providing a boost to the power spectrum on all scales by a constant factor $K_m = (1 + f \mu^2)^2$, with $K$ representing the Kaiser contribution. In this work, we focus on the monopole of the 3D power spectrum, which is given by
\begin{equation} \label{eq:poles}
    P_{\ell} (k) = \frac{2 \ell + 1}{2} \int_{-1}^1 d\mu P(k, \mu) \mathcal{L}_\ell (\mu) \,,
\end{equation}
with $\ell = 0$. The Kaiser contribution to the monopole of the matter power spectrum is $1 + 2/3 f + 1/5 f^2$.

We can obtain the Kaiser term for tracer auto and cross-power spectra by substituting the tracer bias (Equation~\ref{eq:bias}) into the continuity equation $\delta_{m, L} + \nabla \cdot \boldsymbol{v}_L = 0$, yielding $\delta_{t, L} / b_t + \nabla \cdot \boldsymbol{v}_L = 0$. The linear tracer velocity field is assumed to be the same as all matter since both are subject to the same potentials \citep{fry_evolution_1996, tegmark_power_2002, desjacques_modeling_2010}. This results in the following expressions for the Kaiser effect of the auto and cross-power spectra, respectively:
\begin{align}
    K_i &= 1 + \frac{2f}{3 b_i} + \frac{f^2}{5 b_i^2} \,,\label{eq:kaiser_auto} \\ 
    K_{i-j} &= 1 + \frac{f}{3 b_i r_{i-j}} + \frac{f}{3 b_j r_{i-j}} + \frac{f^2}{5 b_i b_j r_{i-j}}\,. 
    \label{eq:kaiser_x}
\end{align}
We strictly use constant bias values since Equations~\ref{eq:kaiser_auto}-\ref{eq:kaiser_x} are only valid when both the velocity and density fields are linear. Of the two fields, the velocity field becomes nonlinear at larger scales because of its higher sensitivity to tidal effects \citep{scoccimarro_redshift-space_2004, jennings_modelling_2011, reid_towards_2011, chan_gravity_2012, howlett_cosmological_2017}. Tracer biases, conversely, are relatively insensitive to the velocity field, typically remaining constant in the linear regime for the density field \citep{seljak_large-scale_2004, jullo_cosmos_2012}. Thus, we can assume the bias is constant when using Equations~\ref{eq:kaiser_auto}-\ref{eq:kaiser_x}. We show the contribution of the Kaiser term to the monopole of the \hi-galaxy cross-power spectrum with the red line in Fig.~\ref{fig:demo}, which comprises a vertical translation of the power spectrum on all scales.

On smaller scales, the random motion of matter within virialized structures stretches the apparent clustering within halos along the line-of-sight. This phenomenon is known as the fingers-of-God (FoG) effect \citep{jackson_critique_1972} and suppresses redshift-space clustering on small scales. The FoG effect is typically modeled with a damping term that can take a variety of different forms. We follow the analysis of \citet{sarkar_modelling_2018}, who tested some common damping profiles for \hi and found a small preference for 
\begin{equation} \label{eq:fog} % this is not squared, since it is squared in the power spectrum equation.
    \dfog{} (k, \mu, \sigma_i) = \frac{1}{1 + \frac{1}{2}k^2 \mu^2 \sigma_i^2} \,,
\end{equation}
where $\sigma_i$ is the pairwise velocity dispersion \citep{peacock_reconstructing_1994, park_power_1994, sheth_linear_2001}. We emphasize that Equation~\ref{eq:fog} is purely phenomenological -- $\sigma_i$ is determined by fitting to the power spectrum. Furthermore, Equation~\ref{eq:fog} assumes that $\sigma_i$ is scale-independent, which is known to be false \citep{scoccimarro_redshift-space_2004, slosar_pairwise_2006, loveday_galaxy_2018} although perhaps justified if $\sigma_i$ is weakly scale-dependent over the scales of interest. The impact of the FoG term can be seen in the yellow line of Fig.~\ref{fig:demo}, suppressing the power spectrum at $k \sim 0.3$ \hmpc with the effect strengthening toward smaller scales.

In many models of RSDs, the FoG and Kaiser effects are handled separately which implicitly assumes that they are decoupled. However, both effects originate from the same line-of-sight displacements and therefore cannot be truly independent of each other \citep{scoccimarro_redshift-space_2004, hikage_fingers--god_2015}. Furthermore, the mapping between real and redshift space in Equation~\ref{eq:rsd_exact} couples the Kaiser and FoG effects \citep{taruya_baryon_2010}. $\Delta u_z$ subtracts out large-scale averages in the velocity field, rendering the associated exponential term sensitive to small-scale effects like the random motion of matter within virialized objects. We speculate on how this RSD treatment affects the performance of the models in Section~\ref{disc:fog}.

\subsection{General model} \label{models:complete}
Combining the ingredients discussed in Sections \ref{model:nonlin}-\ref{models:rsd}, we arrive at the general equation for the \hi auto and \hi-galaxy cross-power spectra models we test in this work:
\begin{equation}
    P^S\shi (k, \mu) =  b\shi^2 (k) K\shi P_m^R(k) D_{\mathrm{FoG}}^2 (k, \mu, \sigma\shi), 
    \label{eq:auto_full}
\end{equation}
\begin{multline}
    P^S_{\mathrm{HI} \times \mathrm{Gal}} (k, \mu) = b\shi (k) b_{\mathrm{Gal}} (k) r_{\mathrm{HI-Gal}} (k) \times \\ K_{\mathrm{HI-Gal}} P_m^R(k) D_{\mathrm{FoG}} (k, \mu, \sigma\shi) D_{\mathrm{FoG}} (k, \mu, \sigma_{\mathrm{Gal}}) \,.
    \label{eq:cross_full}
\end{multline}
Equations~\ref{eq:auto_full}-\ref{eq:cross_full} are broadly representative of the various techniques used to model tracer power spectrum. For example, HOD models typically use the same Kaiser and FoG terms but determine the matter power spectrum and bias via integrals over various occupation properties as a function of halo mass, which requires some additional assumptions \citep[for review, see][]{cooray_halo_2002}. Traditionally, the pairwise velocity dispersion is determined with a fit to the power spectrum, although some works have derived $\sigma\shi$ self-consistently within the halo framework \citep[e.g.,][]{zhang_parameter-free_2020, padmanabhan_2023}. Conversely, LPT approaches modify the Kaiser term to include higher-order components \citep{scoccimarro_redshift-space_2004, percival_testing_2009, taruya_baryon_2010, vlah_lagrangian_2015}. We discuss the potential impact of these higher-order terms in Section~\ref{disc:fog}.

In Fig.~\ref{fig:demo}, we compare the final \hi-galaxy cross-power spectrum model (Equation~\ref{eq:cross_full}, pink line) with all ingredients to the TNG300 values (dotted line). The TNG300 power spectra are calculated by placing the \hi and galaxy distributions into a $800^3$ grid smoothed with a cloud-in-cell assignment scheme. Our results are converged with the sampling of the grid \citep[appendix B in][]{osinga_atomic_2024}. The redshift-space positions of \hi and galaxies are computed using their velocities along an arbitrarily-chosen line of sight. We then compute the power spectra with \textsc{pylians} \citep{villaescusa-navarro_pylians3_2024}. This procedure is used for all power spectra calculated from TNG300 for the remainder of this work. 

The TNG300 and model cross-power spectrum agree reasonably on all scales, effectively by construction since we extract estimates of the ingredients used in the model from TNG300.  Any differences between the two originate from the matter power spectrum and the FoG term. On large scales, the TNG300 values fluctuate around the model values at the largest scales probed by TNG300 ($k \approx 0.05$ \hmpc) due to statistical variance, like the matter power spectrum (Section~\ref{model:nonlin}). On small scales, the model overpredicts the TNG300 cross-power spectrum due to limitations of using a single FoG damping term, which we analyze further in Section~\ref{disc:fog}.

\subsection{Simplifying assumptions} \label{models:assump}

The general models that we test for \hi auto and \hi-galaxy cross-power spectra are described in Equations~\ref{eq:auto_full} and \ref{eq:cross_full}. However, many works choose to make additional explicit assumptions to simplify them further, which is justified if the incurred model errors are negligible compared to the uncertainties in the measurements. Here, we list each assumption and describe how we test its validity.

\textbf{1) Tracer biases and correlation coefficients are constant.} Tracer biases and correlation coefficients are constant in the linear regime, but some studies have shown that nonlinear effects can emerge on surprisingly large scales \citep{umeh_nonlinear_2016, osinga_atomic_2024}. These nonlinearities are often neglected in \hi intensity maps since they are expected to be small compared to measurement uncertainties on large scales. Nonetheless, some works add perturbations to the bias to incorporate some nonlinear terms, like $b(k) \approx b_0 + b_1 k$ \citep[e.g.,][]{mcdonald_clustering_2009,saito_understanding_2014,springel_first_2018}. In our work, we test the two extremes to capture the full range of behavior: $b_c$, a constant bias, and $b (k)$, a completely scale-dependent bias.

\textbf{2) Fingers-of-God can be neglected.}
Since the FoG effect is expected to be a small-scale effect, some large-scale measurements are interpreted with models that exclude the FoG damping term altogether \citepalias[e.g.,][]{wolz_h_2022, cunnington_h_2022}. We therefore compare models with and without an FoG damping term.

\textbf{3) The tracer Kaiser term is approximately the matter Kaiser term.}
Current measurements of the \hi-galaxy cross-power spectra do not yet have the signal-to-noise to extract the quadrupole ($\ell = 2$, Equation~\ref{eq:poles}). Consequently, \citetalias{wolz_h_2022}, \citetalias{cunnington_h_2022}, and \citetalias{Carucci24} use the matter Kaiser term $K_m = (1 + f\mu^2)^2$ instead of the tracer Kaiser term from Equation~\ref{eq:kaiser_x} (see cited works for further explanation). We only test models using the matter Kaiser terms for the \hi-galaxy cross-power spectra to probe the impact of this assumption.

\section{Testing the model ingredients within IllustrisTNG} \label{sec:ing}

In Section~\ref{sec:models}, we described the ingredients used to model the \hi-galaxy cross-power spectra, provided examples of the impact of each ingredient with Fig.~\ref{fig:demo}, and described some common simplifying assumptions we intend to test. In this section, we describe how we extract each ingredient from simulations and evaluate the validity of common assumptions associated with each ingredient, preparing for Section~\ref{sec:perf} in which we test the performance of the models as a whole.

\subsection{Bias} \label{ing:bias}

\begin{figure*}
    \centering
    \includegraphics[width = \linewidth]{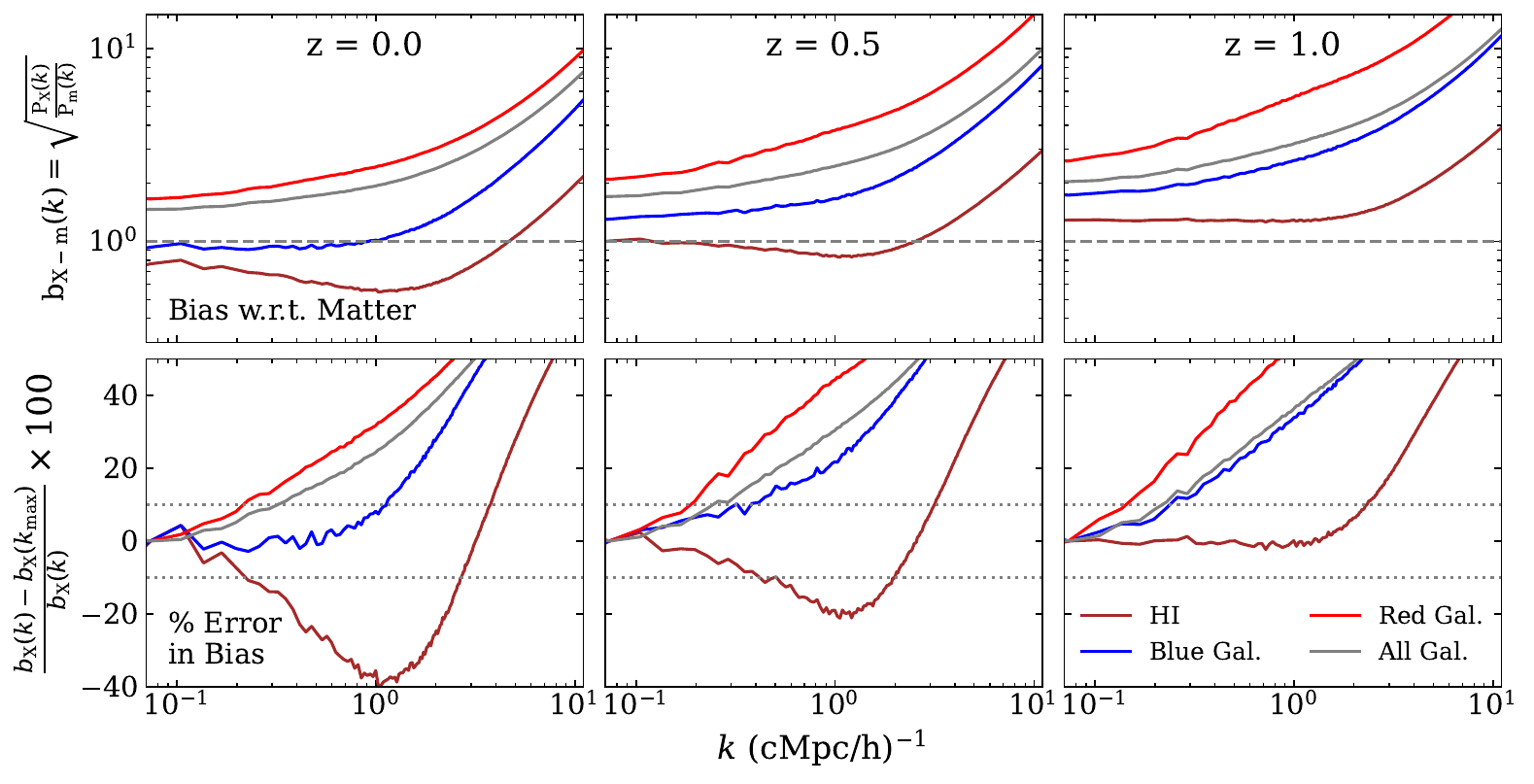}
    \caption{\textit{Top:} Bias with respect to matter for various baryonic tracers. Blue (blue), red (red), and all-galaxies (gray) are shown with \hi (brown) at $z = 0$ (left), $z = 0.5$ (center), and $z = 1$ (right). Large-scale bias values provided in Table~\ref{tab:bias}. \textit{Bottom:} The percentage error caused by assuming a constant bias with gray dotted lines at $\pm 10\%$. The bias for all baryonic tracers at most redshifts differ from the assumed constant value by $\geq 10\%$ by $k \sim 0.2-0.3$ \hmpc. The exceptions are the $z = 0$ blue galaxy bias and $z = 1$ \hi bias, which arise from their low occupation of massive halos offsetting nonlinearities.}
    \label{fig:bias}
\end{figure*}

\begin{figure}
    \centering
    \includegraphics[width = \linewidth]{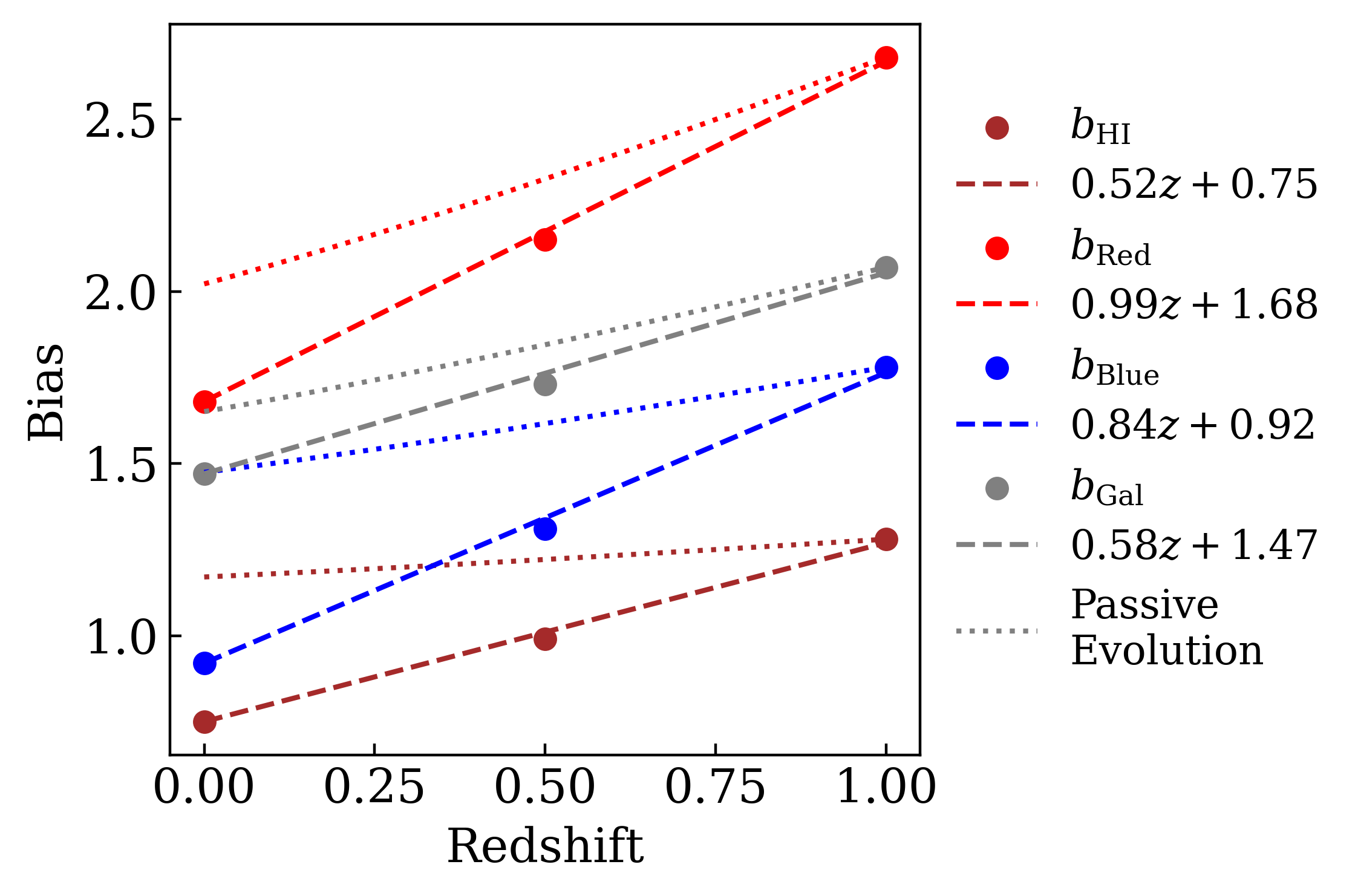}
    \caption{The redshift evolution of the bias for each baryonic tracer, with linear fits (dashed lines) to the data points to help visualize the trends. The precise values presented in these fits should be taken with caution with only three data points. Dotted lines show the expected bias evolution of a population only subject to gravitational effects \citep{fry_evolution_1996}. Tracers associated with star-formation (red, blue, \hi) deviate from the passive evolution more than the all-galaxy bias, suggesting that quenching processes play a significant role in how they trace the matter distribution.}
    \label{fig:evo}
\end{figure}

The \hi and galaxy biases from TNG300 are shown in the top row of Fig.~\ref{fig:bias}. The bias scale-dependencies are influenced largely by two factors: the halo bias and halo occupation \citep{chen_extracting_2021, wang_breakdown_2021}. As explained in Section~\ref{sec:models}, nearly all \hi and galaxies reside in halos and thus reflect the shape of the halo bias. The second factor, halo occupation, modulates this behavior based on how each tracer occupies halos as a function of halo mass. Larger halos are known to cluster more strongly than smaller ones, leading to greater biases and nonlinearities both for them and the tracers that they tend to host \citep{sheth_large-scale_1999,cooray_halo_2002}. For example, red galaxies are known to favor older and massive halos, resulting in a greater amplitude and a sharper slope in the bias at all redshifts \citep{kaiser_spatial_1984, gao_age_2005, zehavi_galaxy_2011}. Conversely, blue galaxies and \hi inhabit smaller, more isolated halos \citep{papastergis_clustering_2013}, resulting in smaller biases and nonlinearities \citep{wang_cross-correlation_2007}. The remaining tracer bias in Fig.~\ref{fig:bias}, all-galaxies, falls between that of blue and red galaxies, as expected from the combination of the two color subsamples.

%Tracer biases are affected by complex nonlinearities that are challenging to include in large-scale structure models. Consequently, most models ignore nonlinearities and assume the bias is constant, which is justified if the error between the constant and actual bias is negligible compared to measurement uncertainties. In this section, we first examine the nonlinear effects that shape the bias of each tracer at $z = 0, 0.5$ and 1, which are presented in the top row of Fig.~\ref{fig:bias}. Afterward, we compare the actual bias to the assumed constant bias in the bottom row of Fig.~\ref{fig:bias} and assess its validity using estimates of the measurement uncertainties from recent observations.

The tendency of \hi and blue galaxies to occupy small halos opposes the scale-dependent contributions from the halo bias, which generates some of the unique bias shapes in Fig.~\ref{fig:bias}. For example, the $z = 1$ \hi bias and $z = 0$ blue galaxy bias are constant to surprisingly small scales because the halo bias and occupation effects coincidentally offset. The occupation effect grows with time, even becoming dominant at 0.1 $\lesssim k \lesssim 1$ \hmpc in the \hi bias at $z < 1$ and creating the ``dip'' in the \hi bias at those scales. These \hi bias shapes agree with other theoretical \citep{penin_scale-dependent_2018}, computational \citep{wolz_intensity_2016, spinelli_atomic_2020, wang_breakdown_2021}, and observational \citep[$z = 0$,][]{anderson_low-amplitude_2018} works on \hi distributions. We note however that the amplitude of the \hi bias can vary substantially between the different works (further discussed in Section~\ref{disc:sims}).

The scale-dependency of each tracer bias determines the scales at which a constant bias can be assumed without inheriting significant errors, shown in the bottom row of Fig.~\ref{fig:bias}. We define the ``valid'' scales to have errors smaller than 10\% (gray dotted lines), with \kten representing the scale at which the error crosses this threshold. The 10\% threshold corresponds to $\sim 1/10$ of the measurement uncertainties from \citetalias{cunnington_h_2022}, such that \kten roughly coincides with the scale at which differences between the scale-dependent and constant bias are no longer negligible. We note that the improved handling of systematics in \citetalias{Carucci24} have already reduced these uncertainties, making the theoretical errors more impactful. More faithful comparisons require instrument-by-instrument mock observations --- we address these in Section~\ref{omega:revisions}.

We quantify the scale-dependency of each tracer by measuring deviations from their large-scale values, assumed to be constant. The constant biases are estimated by averaging over 0.043 \hmpc $\leq k \leq$ 0.105 \hmpc ($k_{\mathrm{eff}}$ = 0.08 \hmpc), mitigating the impact of the small number of samples in the largest $k$-modes. These large-scale biases and their corresponding \kten values are given in Table~\ref{tab:bias}. The bottom row of Fig.~\ref{fig:bias} shows that red galaxy bias deviates the most from its large-scale limit, while blue galaxies and \hi deviate the least. This result follows from the analysis of how the halo bias and occupation effects manifest in each tracer bias --- namely, each effect compounds in red galaxies and offset in \hi and blue galaxies.

After studying the scale-dependence of the tracer biases, we now turn to their redshift evolution, which can provide insight into the role that gravity plays in tracer distributions. A population in which gravity solely governs the evolution of its distribution is called ``passively evolving'' with its bias relaxing toward unity:
\begin{equation} \label{eq:pass_evo}
    b_0 = \frac{b (z) + z}{1 + z} \,.
\end{equation}
Significant deviations from passive evolution suggest the influence of additional processes on the tracer's clustering beyond gravity, such as changes in number density through mergers or satellite disruption \citep{bell_nearly_2004, guo_clustering_2013, skibba_primus_2014}. We compare the redshift evolution of the various tracer biases to the expected passive evolution in Fig.~\ref{fig:evo}. We fit a linear model to the three data points, with the slope as the only free parameter and assuming the $z=0$ value as the intercept. 

\begin{deluxetable}{ccccccccc}
    \tablecaption{\hi and Galaxy Biases}
    \label{tab:bias}
    \tablewidth{0pt}
    \tablehead{
    \colhead{} & \multicolumn{2}{c}{\hi} & \multicolumn{2}{c}{Blue Gal.} & \multicolumn{2}{c}{Red Gal.} & \multicolumn{2}{c}{All Gal.} \\
    \cline{2-9}
    \colhead{$z$} & \colhead{$b_c$} & \colhead{$\kten$} & \colhead{$b_c$} & \colhead{$\kten$} & \colhead{$b_c$} & \colhead{$\kten$} & \colhead{$b_c$} & \colhead{$\kten$}
    }
    \startdata
    0   & 0.75 & 0.22 & 0.92 & 1.15 & 1.68 & 0.23 & 1.47 & 0.35 \\
    0.5 & 0.99 & 0.41 & 1.31 & 0.32 & 2.15 & 0.20 & 1.73 & 0.26 \\
    1   & 1.28 & 2.22 & 1.78 & 0.26 & 2.68 & 0.17 & 2.07 & 0.23 \\
    \enddata
    \tablecomments{Bias values measured from TNG300, averaged across 0.043 $\leq k \leq 0.105$ \hmpc to reduce noise, although the averaged values agree closely with the large-scale values anyway ($\lesssim 1.5\%$). \kten (\hmpc) represents the scale at which these bias values deviate by $10\%$ from the actual scale-dependent bias. This threshold was chosen to represent when model errors become significant to measurement uncertainties from \citetalias{cunnington_h_2022}. The reported \kten values should be interpreted as an optimistic estimate for this threshold.}
\end{deluxetable}

The key conclusion from Fig.~\ref{fig:evo} is that tracers sensitive to star-formation deviate more from passive evolution than tracers independent of star-formation. This trend arises primarily from changes in how star-formation dependent tracers occupy halos. At an earlier redshift, the star-forming galaxies that occupy the most massive, clustered halos also tend to be closest to quenching. Once these galaxies quench at a later redshift, the star-forming galaxies lose their most clustered component, inducing weaker clustering on average. However, these recently-quenched galaxies join the quiescent population as their least massive, clustered members, also suppressing clustering. Therefore, galaxy quenching reduces the clustering of \textit{all} tracers dependent on star-formation, like \hi and blue and red galaxies \citep[for more details, see][]{osinga_atomic_2024}. The stronger role of galaxy formation physics on the distribution of \hi and blue galaxies at low redshifts suggests that they may be poorer tracers of the matter distribution than a star-formation independent population.

\begin{figure*}
    \centering
    \includegraphics[width=\linewidth]{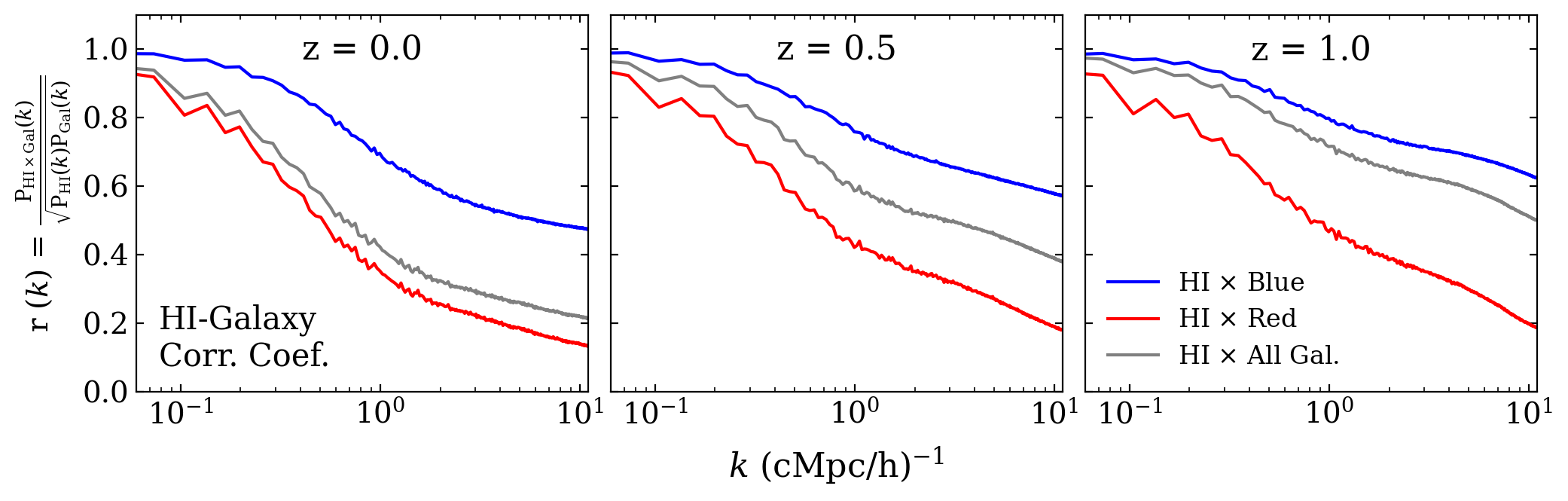}
    \caption{Correlation coefficients of \hib, \hir, and \hia for $z = 0$ (left), $z = 0.5$ (center), and $z = 1$ (right). Large-scale values are presented in Table~\ref{tab:cc}. \hi correlates more strongly with blue galaxies than red, leading to a larger amplitude and less scale-independence in \hib. \hia lies between \hib and \hir, representing an effective average of its two subpopulations. The slope of each correlation coefficient becomes steeper with time as more galaxies quench.}
    \label{fig:cc}
\end{figure*}

\begin{deluxetable}{ccccccc}
    \tablecaption{\hi-Galaxy Correlation Coefficients}
    \label{tab:cc}
    \tablewidth{0pt}
    \tablehead{
    \colhead{} & \multicolumn{2}{c}{\hib} & \multicolumn{2}{c}{\hir} & \multicolumn{2}{c}{\hia} \\
    \cline{2-7}
    \colhead{} & \colhead{$r_c$} & \colhead{$\kten$} & \colhead{$r_c$} & \colhead{$\kten$} & \colhead{$r_c$} & \colhead{$\kten$}
    }
    \startdata
    $z = 0$   & 0.98 & 0.35 & 0.86 & 0.17 & 0.90 & 0.17 \\
    $z = 0.5$ & 0.98 & 0.41 & 0.88 & 0.23 & 0.93 & 0.26 \\
    $z = 1$   & 0.98 & 0.44 & 0.87 & 0.23 & 0.95 & 0.32 \\
    \enddata
    \tablecomments{Large-scale correlation coefficients for \hib, \hir, and \hia and the scale at which the actual correlation coefficient value differs from the large-scale value by more than $10\%$ ($\kten$, in \hmpc). The correlation coefficient values are averaged over 0.043 \hmpc $\leq k \leq 0.105$ \hmpc to reduce noise. \hib and \hir evolve negligibly with redshift at large scales. Conversely, all correlation coefficients decrease on small scales with time, increasing the \kten values.}
\end{deluxetable}

We note that the biases in this section include shot noise to follow previous literature, although the shot-noise-independent definition is theoretically preferred. We compare the two bias definitions in Appendix \ref{app:SN} --- we emphasize that both definitions result in the same errors and only differ in how the errors are interpreted. We also note that bias values can be sensitive to resolution, although our estimates of the model errors are robust to them (see Section~\ref{disc:sims} and Appendix~\ref{app:resolution}). 

\subsection{Correlation coefficients} \label{ing:cc}

Similar to the bias studied in the previous section, the correlation coefficients presented in Fig.~\ref{fig:cc} capture complex effects on small scales that many models choose to neglect. Consequently, we adopt a similar analysis in this section, starting with the factors that shape the correlation coefficients before concluding with estimates of the scales at which an assumed constant correlation coefficient is valid.

Fig.~\ref{fig:cc} shows that \hi and galaxies correlate strongly on large scales, as expected since all populations trace the matter distribution. On small scales, however, stochasticity, nonlinearities, and baryonic physics decouple the tracers, reducing the correlation coefficient.

% However, the scale-dependencies of the different correlation coefficients differ in detail. For instance, the slope of the \hir coefficient is visibly sharper than \hia or \hib between 0.1 \hmpc $\lesssim k \lesssim$ 1 \hmpc. When comparing the coefficients across redshifts, it is also clear that the slopes of the later redshifts are steeper than the earlier ones.  

The small-scale decoupling is weakest for \hi and blue galaxies. Star-forming blue galaxies tend to be gas-rich and occupy the same regions of space as \hi, and thus strongly correlate even on small scales \citep{bigiel_star_2008, huang_arecibo_2012, kauffmann_re-examination_2013, papastergis_clustering_2013, hearin_physical_2016, anderson_low-amplitude_2018}. Conversely, red galaxies weakly correlate with \hi because they tend to inhabit gas-poor regions. \hia resides between \hir and \hib, effectively representing their weighted average. A larger fraction of galaxies are quenched at later redshifts such that \hia approaches \hir with time \citep{nelson_first_2018}. This quenching effect does not manifest in the large-scale values of \hib and \hir, although they are reduced on small scales. Table~\ref{tab:cc} quantifies both the redshift and color trends with estimates for the constant correlation coefficients and \kten (definition in Section~\ref{ing:bias}). %\hib retains a larger coefficient and \kten values than \hir and \hia, as expected from the color trends in Fig.~\ref{fig:cc}. The redshift evolution is seen in the \hia values, which begin near \hib at $z = 1$ before approaching \hir at later times. All \kten values decrease with time due to the continued galaxy quenching, although \hib again is the most scale-independent.

Since \hib is less scale-dependent than \hir and \hia, one would na\"ively conclude that the \hib cross-power spectrum is easier to model with constant bias and correlation coefficients. However, the scale-dependence of the bias \textit{opposes} that of the correlation coefficient; the bias (with some exceptions, see Section~\ref{ing:bias}) increases on small scales, while the correlation coefficient decreases. Thus, the product $b\shi b_{\mathrm{Gal}} r_{\mathrm{HI-Gal}}$ in the cross-power spectrum (Equation~\ref{eq:cross_full}) may appear to be more or less scale-independent than each term individually (Appendix~\ref{app:ax_compare}).

\subsection{Redshift-space distortions} \label{ing:rsd}

\begin{deluxetable*}{c|c|c|c|c|c}

    \label{tab:sigma}
    \tablewidth{0pt}
    \tablecaption{Pairwise Velocity Dispersions}
    \tablehead{
    \colhead{$\sigma_p$ (cMpc/$h)^2[\chi^2 / \mathrm{dof}]$} & \colhead{Matter} & \colhead{\hi} & \colhead{Blue Gal.} & \colhead{Red Gal.} & \colhead{All Gal.}
    }
    \startdata
    $z = 0$   & 4.02 [55.0]  & 2.40 [9.72]  & 2.40 [22.0]  & 3.59 [59.2]  & 3.45 [40.2] \\
    $z = 0.5$ & 3.75 [9.57]  & 3.12 [8.74]  & 2.95 [34.5]  & 2.93 [56.2]  & 3.02 [20.3] \\
    $z = 1$   & 2.92 [5.87]  & 2.70 [4.73]  & 2.18 [14.3]  & 2.11 [65.1]  & 2.27 [16.0] \\
    \enddata
    \tablecomments{Pairwise velocity dispersion (PVD) values retrieved from fits of the FoG damping term to the 2D power spectrum presented in Fig.~\ref{fig:2Dpk}. We fit down to $k \leq 1$ \hmpc, which limits shot noise contributions. The $\chi^2$ values are calculated over the same scale regime. The galaxy populations generally have large residuals --- we tested different $k$ limits and found no significant improvement in $\chi^2$, although the PVD values change by $\lesssim 0.1$. We attribute the poor fits to absent nonlinearities and scale-dependent effects in the FoG term, which has been found in other work \citep{ando_redshift_2019}.}
\end{deluxetable*}

% Open question about this section: why are all of the RSD errors seemingly offset by similar amounts in the same direction? I would have thought that the errors would be more spread out, as in both above and below the line. I originally thought maybe this was related to some shot noise contribution, but the fact that the matter error trends the same tells me its not. Also, I think that the shot noise causes errors in the other direction - the model will underestimate the clustering (so below unity), so this can't be from shot noise. 
\begin{figure*}
    \centering
    \includegraphics[width=\linewidth]{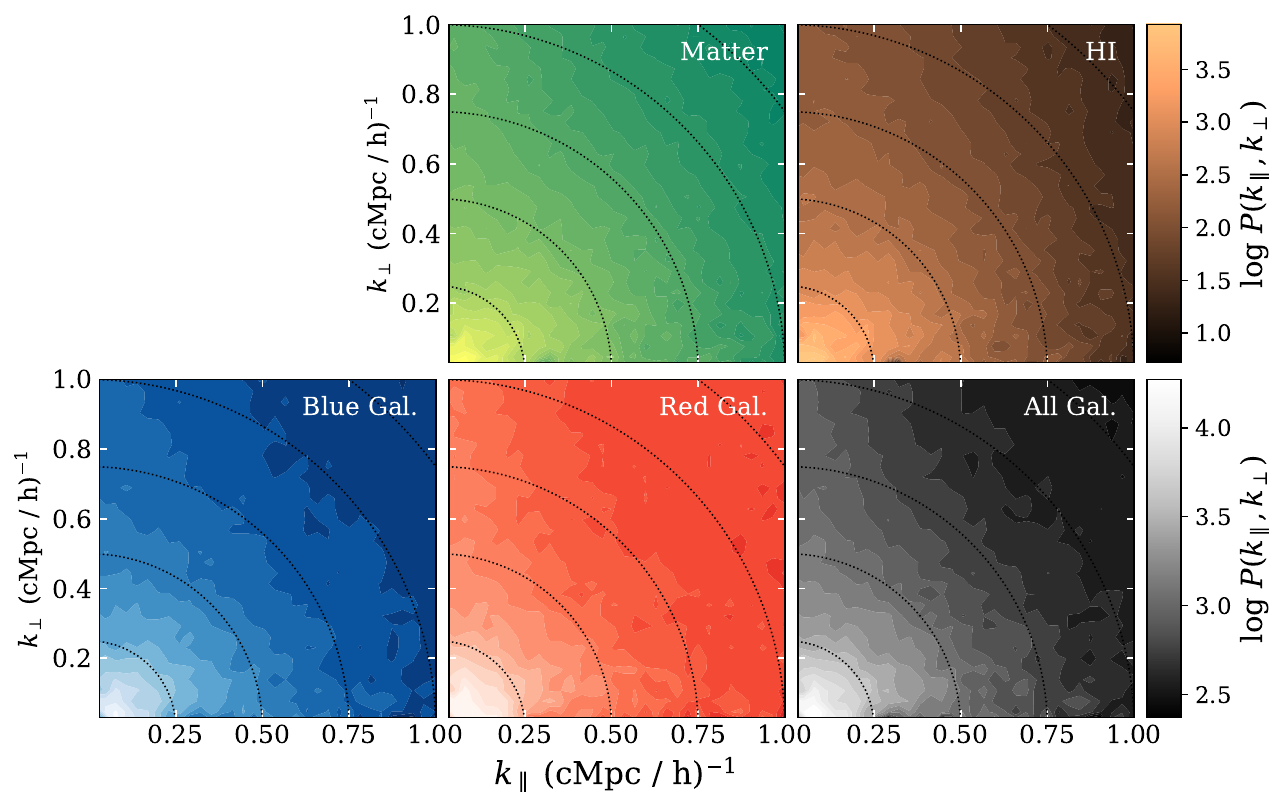}
    \caption{The 2D auto power spectra of all matter (top center), \hi (top right), and blue (bottom left), red (bottom center), and all-galaxies (bottom right) at $z=1$. Black dotted lines are concentric circles to help visualize the FoG effect, which vertically warps the isopower contours. This warping is strongest in red galaxies as compared to the other galaxy types, agreeing with observations \citep{madgwick_2df_2003}. We find that \hi \textit{appears} to have stronger FoG suppression than matter, as found in \citetalias{villaescusa-navarro_ingredients_2018}, although their relative strengths depend on scale (see text). Power spectra are slightly smoothed by a Gaussian filter ($\sigma = 0.5$) to see the contour lines more clearly, but the smoothed values are not used for the determination of the pairwise velocity dispersion and do not otherwise impact our conclusions.}
    \label{fig:2Dpk}
\end{figure*}

\begin{figure}
    \centering
    \includegraphics[width=0.85\linewidth]{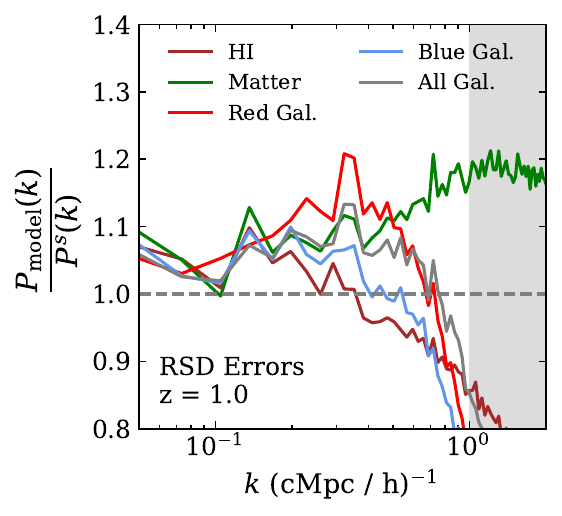}
    \caption{Errors arising from how RSDs are modeled for each distribution in Fig.~\ref{fig:2Dpk}, expressed as a ratio of the model power spectra ($P_{\mathrm{model}}$) over the corresponding redshift-space power spectra from IllustrisTNG. $P_{\mathrm{model}}$ is obtained by substituting the fiducial $\sigma_p$ and $b$ values for each distribution into Equation~\ref{eq:rsd_no_nl}, which is then integrated according to Equation~\ref{eq:fog_fit_tests}. The gray shaded region denotes scales that were not included in the $\sigma_p$ fit or $\chi^2$ analysis (Table~\ref{tab:sigma}). The errors demonstrate that the models systematically overestimate the actual power spectrum until $k \sim 1$ \hmpc.    %and the corresponding actual auto power spectrum. We obtain the 3D power spectrum by integrating  obtain the $\sigma_p$ values for each tracer from fits of Equation~\ref{eq:rsd_no_nl} to the 2D power spectrum (Fig.~\ref{fig:2Dpk}). We then integrate Ratios of the integrated monopole (Equation~\ref{eq:fog_fit_tests}) over the actual redshift-space power spectrum, with each color corresponding to the population's color map in the respective panel. We find that the modeled 2D power spectra do not accurately replicate the 1D redshift-space power spectra, which we attribute to how RSDs are handled in phenomenological models.
    }
    \label{fig:RSD_errors}
\end{figure}

Although all redshift-space distortions (RSDs) arise from line-of-sight velocities displacing the apparent positions of tracers, models often treat them as two separate contributions (see Section~\ref{sec:models}): the Kaiser and fingers-of-God (FoG) effect. A useful way to visualize these RSDs is to plot the power spectrum as a function of wavenumbers parallel ($k_\parallel$) and perpendicular ($k_\perp$) to the line of sight. In this section, we analyze how RSDs manifest in the matter, \hi, and blue, red, and all-galaxy 2D power spectra (Fig.~\ref{fig:2Dpk}) and quantify them in Table~\ref{tab:sigma}. 

The Kaiser effect boosts the clustering along $k_\parallel$, squeezing the isopower contours in the bottom left corner of each panel from Fig.~\ref{fig:2Dpk} and creating horizontally-aligned ellipses. This effect's magnitude appears similar amongst the various tracers in Fig.~\ref{fig:2Dpk}, but each Kaiser term should be inversely related to the bias of the tracer (Equation~\ref{eq:kaiser_auto}). We will discuss the Kaiser effect in more detail in Section~\ref{disc:fog} --- for the remainder of this section, we focus on the FoG effect.

The FoG effect suppresses clustering at small scales, elongating the contour lines in the top left corners of each panel in Fig.~\ref{fig:2Dpk}. Amongst the galactic populations in the bottom row, the contour lines in the red galaxy panel are warped the furthest vertically, agreeing qualitatively with observations \citep{madgwick_2df_2003}. The red galaxies' large FoG suppression originates from their tendency to occupy massive halos with large velocity dispersions \citep{zehavi_galaxy_2011}. Conversely, blue galaxies occupy smaller halos, resulting in a weaker FoG effect. The all-galaxies' (bottom right panel) FoG strength falls between blue and red galaxies', as expected for a combined sample.
Of the two remaining populations in the top row of Fig.~\ref{fig:2Dpk}, the FoG damping in \hi \textit{appears} to be stronger than that of matter, as seen from the more vertical contour lines in the panel's top right corner. This agrees with conclusions from \citetalias{villaescusa-navarro_ingredients_2018}, but the following discussion about Table~\ref{tab:sigma} will demonstrate that this finding does not hold quantitatively.
%However, the conclusions drawn from the contour lines contradict the larger ``pairwise velocity dispersion'' (PVD, see Section~\ref{sec:models}) values for matter in Table~\ref{tab:sigma}, which would imply that matter has the stronger FoG.

We measure the strength of the FoG effect with the ``pairwise velocity dispersion'' (PVD, see Section~\ref{sec:models}). This parameter is difficult to measure directly \citep[e.g.,][]{loveday_galaxy_2018}, particularly in intensity maps which do not observe point sources. Instead, the PVD is usually left as a free parameter to marginalize over \citep[e.g.,][]{gil-marin_completed_2020, demattia_completed_2021}. Following \citetalias{villaescusa-navarro_ingredients_2018}, we extract the PVD (denoted $\sigma_p$) by fitting the following equation to the 2D power spectra of the population $i$ on scales $k \leq 1$ \hmpc:
\begin{equation} \label{eq:rsd_no_nl}
    P^S_i (k, \mu) = \left(1 + \frac{f \mu^2}{b_i}\right)^2 P^R_i (k) \left( \frac{1}{1 + \frac{1}{2} k^2 \mu^2 \sigma_p^2}\right)^2 \,,
\end{equation}
where $\mu = \hat{k_\parallel} \cdot \hat{k}$. We present the resulting PVD values in Table~\ref{tab:sigma}.

Most PVDs increase with time, as expected from the growth of halos bolstering their velocity dispersions. This manifests in the matter, red galaxy and all-galaxy PVDs, with the red galaxy PVDs growing rapidly enough to overtake all-galaxies by $z = 0$. Blue galaxies and \hi, however, do not evolve as expected between $z = 0.5$ and 0, due to their decreasing occupation of the most massive halos which typically have the strongest dispersions (see Section~\ref{ing:cc}). This occupation effect should also manifest between $z = 1$ and $z = 0.5$, but the number of quenching galaxies forms a smaller proportion of the blue and \hi-rich population, mitigating its strength.

However, the conclusions from Fig.~\ref{fig:2Dpk} about the relative strengths of FoG damping amongst the populations are not supported by Table~\ref{tab:sigma}. For example, at $z = 1$, red galaxies possess the smallest PVD despite their apparently large FoG suppression in Fig.~\ref{fig:2Dpk}. Out of the galaxy populations at this redshift, the all-galaxies population actually exhibits the largest PVD. Furthermore, despite the more vertical contours in \hi, we find that the \hi PVD is smaller than the matter PVD, conflicting with conclusions from \citetalias{villaescusa-navarro_ingredients_2018}.

Some of the tension between Fig.~\ref{fig:2Dpk} and Table~\ref{tab:sigma} can be credited to the poor fits, which we quantify with the $\chi^2$ values in Table~\ref{tab:sigma}. \hi and matter generally maintain small $\chi^2$ values, but the galaxy populations all contain large values. To ensure that the poor fits originate from the model rather than our methodology, we tested the sensitivity of the $\chi^2$ to the scales we included in our fits ($k \leq 1$ \hmpc). All tests produced larger $\chi^2$, although we found that the scale regime could change the PVD values by $\sim$ 0.1. PVD is a scale-dependent quantity (see Section~\ref{models:assump}), so the sensitivity to scale is expected. These tests suggest the poor fits stem from representing RSDs with Equation~\ref{eq:fog} for blue and red galaxy distributions, which agrees with results from other works \citep[e.g.,][]{hikage_fingers--god_2015, orsi_impact_2018_fix}. We also tested a single Lorentzian \citep[following][]{sarkar_modelling_2018} to represent the FoG term for the \hi auto power spectrum (as compared to the Lorentzian squared from Equation~\ref{eq:auto_full}), but found no improvement in performance, agreeing with conclusions from \citetalias{villaescusa-navarro_ingredients_2018}.

% reminder for self - don't describe the roles of FoG/Kaiser terms in the error here, that discussion is reserved for the discussion section.
To understand how these fits shape the model errors examined in the following section, we compare the integrated 2D power spectrum model, defined as 
\begin{equation}\label{eq:fog_fit_tests}
    P_{\mathrm{1D}, i} (k) = \frac{1}{2} \int_{-1}^1 P_i^S (k, \mu) d\mu \,,
\end{equation}
to the actual redshift-space power spectrum for each population in Fig.~\ref{fig:RSD_errors} (additional redshifts provided in the online figures). Generally, we find that the integrated 2D power spectra overestimate the actual redshift-space power spectra for each population, and particularly for galaxies. The \hi ratio falls below unity at $k \sim 0.4$ \hmpc, although all of the galaxy tracers follow suit at smaller scales ($k \sim 0.9$ \hmpc). We attribute this feature in the galaxy ratios to the shot noise floor that they reach around that scale, visible in their redshift-space power spectra (online figures). We limit the impact of shot noise by restricting our fits to $k \leq 1$ \hmpc, but as previously mentioned, the fits do not improve with more or less conservative $k$ limits.

In summary, we provide the 2D power spectra for matter and each tracer in Fig.~\ref{fig:2Dpk}, and fit PVD values to them using the phenomenological model from Equation~\ref{eq:rsd_no_nl}. However, we find that these fits misrepresent the 1D power spectrum when integrated, which may arise from the particular functional form in Equation~\ref{eq:fog} as we speculate in Section~\ref{disc:fog}. For now, we will broadly refer to the differences shown in Fig.~\ref{fig:RSD_errors} as ``RSD errors''. In the following sections, we do not display error bars that arise from uncertainties in the PVD values because they are not statistical in nature (Fig.~\ref{fig:RSD_errors}).

\section{Testing the model predictions} \label{sec:perf}

In Section \ref{sec:ing}, we discussed each ingredient of the general model in Equations~\ref{eq:auto_full}-\ref{eq:cross_full} for \hi auto and cross-power spectra. The goal of this section is to understand the aggregate effect of each ingredient on the model's performance by comparing their predictions to the simulated power spectra, seeking to answer the question ``Given the \hi distribution in IllustrisTNG, how accurately could these models extract cosmological parameters from a perfectly measured power spectrum?'' We know that any deviations between the models and simulated power spectra must arise from the combination of these ingredients failing to capture tracer clustering behavior, since each ingredient is extracted directly from the simulation. Moreover, these ingredients will be realistically further degraded by systematics and noise in observations, which are not present in our analyses. As such, we emphasize that \textit{our tests represent the best-case scenario} and stated errors should be taken as lower limits. Qualitatively, the following results should be robust to cosmic variance although second-order effects could change some details \citep[e.g.,][]{li_super-sample_2014}.

\subsection{\hi auto power spectra} \label{perf:auto}

\begin{figure}
    \centering
    \includegraphics[width=\linewidth]{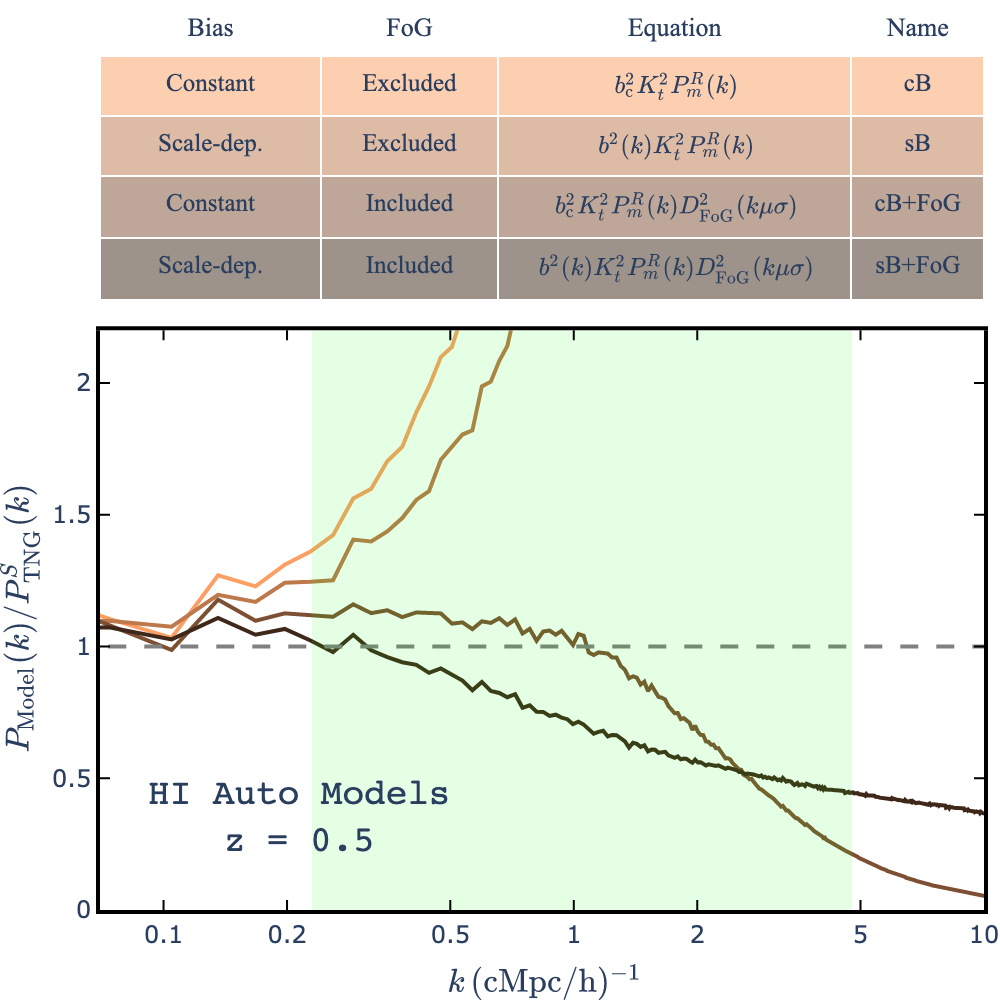}
    \caption{Models of the $z = 0.5$ \hi auto power spectrum, shown as a ratio over the computed values from TNG300. Colors range from lightest to darkest in order of simplest to most complex (see table above). Better performing models remain close to unity to smaller scales. For reference, the green area corresponds to the scale of the MeerKAT observations from \citetalias{paul_first_2023_eprint}. As expected, the more terms we add, the better the model performs, with the FoG term having the strongest impact.}
    \label{fig:auto_mod}
\end{figure}

The \hi auto power spectrum in IllustrisTNG agrees closely with recent observations of the $z \approx 0.5$ \hi auto power spectrum \citep[\citetalias{paul_first_2023_eprint};][]{ osinga_atomic_2024}, presenting a great opportunity to leverage IllustrisTNG to test common models of large-scale \hi distributions. In Fig.~\ref{fig:auto_mod}, we compare four models of varying complexity by showing their ratios over the actual $z = 0.5$ redshift-space \hi auto power spectrum from IllustrisTNG, with darker colors corresponding to more complex models (plots for additional redshifts are provided in Appendix \ref{app:addz}). The details of each prescription are described in the legend, with the background color of each row matching the corresponding power spectrum. The first two columns of the legend indicate whether or not the model assumes a constant bias or neglects the FoG effect. The third column details the equivalent equation, and the fourth column includes a shorthand that we use in the text.

On the largest scales, we expect small differences between the various models since the assumptions made by the simpler prescriptions should be appropriate (see Section~\ref{models:assump}). While all models converge on the largest scales, interestingly they do not converge to the actual redshift-space \hi auto power spectrum, overestimating it by a factor of 1.08. Noise emerging from a small number of large-scale could contribute but is unlikely to account for this difference entirely. We instead attribute the large-scale offset to the Kaiser term since all models err by the same amount regardless of bias or FoG treatment. The Kaiser term neglects nonlinearities in the matter velocity fields, which are known to emerge at $k \sim 0.05$ \hmpc \citep{reid_towards_2011, jennings_modelling_2011, howlett_cosmological_2017}. Despite the departures on the largest scales, all models remain within 10\% of the actual power spectrum at $k \lesssim 0.15$ \hmpc.

The two simplest prescriptions cross the $10\%$ threshold at similar scales, whether a constant (\amA) or scale-dependent (\amB) bias is assumed. The small difference in performance between \amA and \amB, despite the impractically precise description of the \hi bias in \amB, suggests that including non-constant bias terms in \hi auto power spectra models would negligibly improve their performance. The errors in these models worsen to $>25\%$ on the scales relevant to a recent observation of the \hi auto power spectrum, \citetalias{paul_first_2023_eprint}\footnote{We note that \citetalias{paul_first_2023_eprint} remove $k_\parallel \leq 0.3 k_\perp$ modes from their analysis due to instrumentation effects, which should change model performance. As such, Fig.~\ref{fig:auto_mod} should be interpreted as the model errors from a similar survey without this limitation.} (green shaded area). Most of the error from the \amA and \amB models arises from neglecting the FoG term, as shown by the significant improvement in \amC. Despite neglecting non-constant terms in the \hi bias, the \amC error on observation scales remains at $\lesssim 15\%$ until $k \sim 1.5$ \hmpc.

\amD includes a fully scale-dependent bias and FoG term, such that RSD errors (Section~\ref{ing:rsd}) are the only remaining error sources. \amD improves on \amC mostly at large scales, until $k \sim 0.6$ \hmpc where their relative error inverts. \amC can perform better than \amD because it has two sources of error that oppose each other: the overestimated \hi bias and underestimated RSD effects. This phenomenon is unique to \hi, as the other tracers do not possess the intermediate-scale dip that allows a constant bias to overestimate the actual bias (Fig.~\ref{fig:bias}). Other tracers instead always underpredict the bias, which compounds with the RSD error.  

% Assuming a constant \hi bias overestimates the actual value until $k \sim 2$ \hmpc, where the trend inverts and constant bias underestimates the actual value (Fig.~\ref{fig:bias}). On the other hand, the RSD errors underestimate the actual \hi redshift-space power spectrum on $k \gtrsim 0.3$ \hmpc. The overestimated \hi bias and underestimated redshift-space effects reduce the net error in \amC, whereas only the latter source remains in \amD. 

In summary, Fig.~\ref{fig:auto_mod} shows that neglecting the FoG term, even on scales $k \sim 0.1$ \hmpc, causes inaccuracies that dominate other potential error sources, like neglecting non-constant terms in the \hi bias. We also demonstrate that competing error sources can reduce the net error of a simple model, to the point it appears more ``accurate'' than a more complex model. However, even the full prescription incurs an error of $>10\%$ on observationally relevant scales, which would significantly skew the cosmological constraints inferred from \hi auto power spectra.

\subsection{H\textsc{i}-galaxy cross-power spectra} \label{perf:x}

\begin{figure*}
    \centering
    \includegraphics[width=\linewidth]{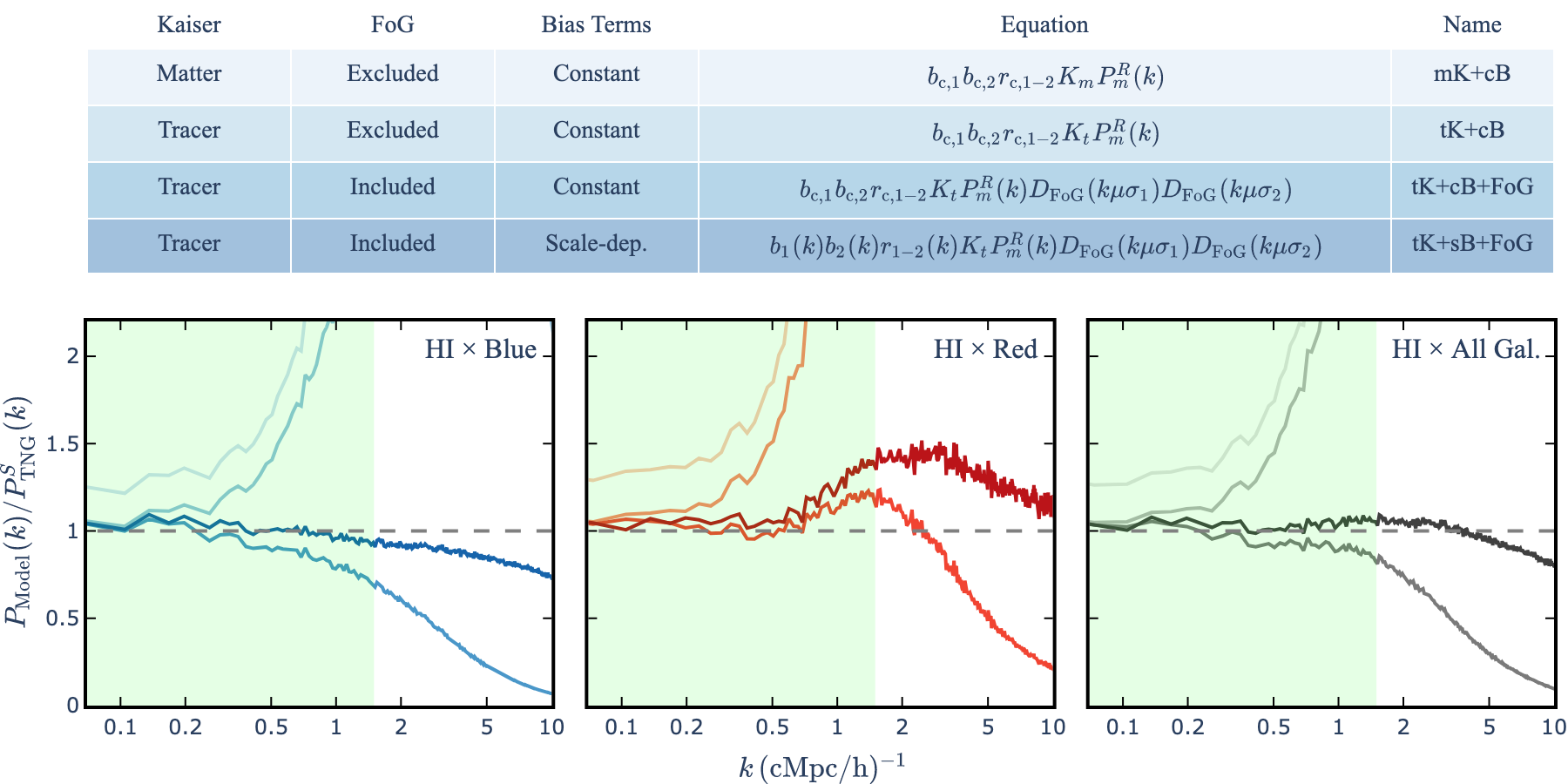}
    \caption{Accuracy of models of \hib (left), \hir (center), and \hia (right) at $z = 1$, compared to IllustrisTNG. Darker colors correspond to more complex models (see legend). Green areas show the typical scales of observations. The simplest prescription represents the model used to interpret recent observations, which errs by $\sim 25\%$ even to the largest scales probed by TNG300. Better performing models stay near unity to smaller scales. As expected, more complex models capture more of the behavior of their respective cross-power spectra. However, even the full prescription for \hir errs by $>50\%$ at $k \sim 1$ \hmpc.}
    \label{fig:xpk_mod}
\end{figure*}

Cross-correlations between \hi and galaxy surveys reduce the systematics and noise present in the \hi auto power spectrum, at the cost of additional complexity (Equations \ref{eq:auto_full}-\ref{eq:cross_full}). In this section, we investigate model errors of \hi-galaxy cross-power spectra in Fig.~\ref{fig:xpk_mod} at $z = 1$, coinciding with most observations \citep{chang_intensity_2010, masui_measurement_2013, wolz_h_2022, 2023_chime_citation}. Additional redshifts are provided in Appendix \ref{app:addz}.

Although both Fig.~\ref{fig:auto_mod} and \ref{fig:xpk_mod} show four models each, the models in the latter differ from the former in two ways. First, recent measurements of the \hi-galaxy cross-power spectrum used the matter Kaiser term \citepalias[Equation~\ref{eq:kaiser_matter}, ][]{wolz_h_2022, cunnington_h_2022, Carucci24}, which we include in Fig.~\ref{fig:xpk_mod}. As a result, the naming convention for cross-power spectra models includes a specification for the Kaiser term, whereas auto power spectra only include the type of bias and FoG terms (see legends of Fig.~\ref{fig:auto_mod} and \ref{fig:xpk_mod}). This additional specification distinguishes the auto and cross-power spectra model shorthands. Second, we remove a cross-power spectra model that uses a scale-dependent bias but neglects the FoG effect (analogous to \amB from Section~\ref{perf:auto}) since we established in the previous section that neglecting FoG is the dominant source of error.

Out of the models in Fig.~\ref{fig:xpk_mod}, we find \xmA incurs the largest error, overpredicting the fiducial power spectra by $\gtrsim 25\%$ even on the largest scales. The error is larger for populations with biases that deviate further from unity like red galaxies, such that the matter (Equation~\ref{eq:kaiser_matter}) and tracer (Equation~\ref{eq:kaiser_x}) Kaiser terms diverge. This difference can be seen in \xmB, which only differs from \xmA in the Kaiser term. Since recent observations interpret their measurements with \xmA models, the large error has important ramifications on their cosmological constraints, which we analyze in Section~\ref{perf:omega}.

Similar to Section~\ref{perf:auto}, we find that including an FoG term substantially enhances model performance even to the largest scales probed by TNG300, as shown in the comparison between \xmB and \xmC. However, the improvement manifests uniquely among the different galaxy populations due to how the non-constant bias terms and RSD errors offset each other. For example, these two error sources coincidentally cancel at $z = 1$ for \hir but do not for \hib or \hia, despite the models for \hib and \hia incurring smaller errors in both biases and RSDs separately (see Sections~\ref{ing:bias} and \ref{ing:rsd}).

\begin{figure*}
    \centering
    \includegraphics[width=\linewidth]{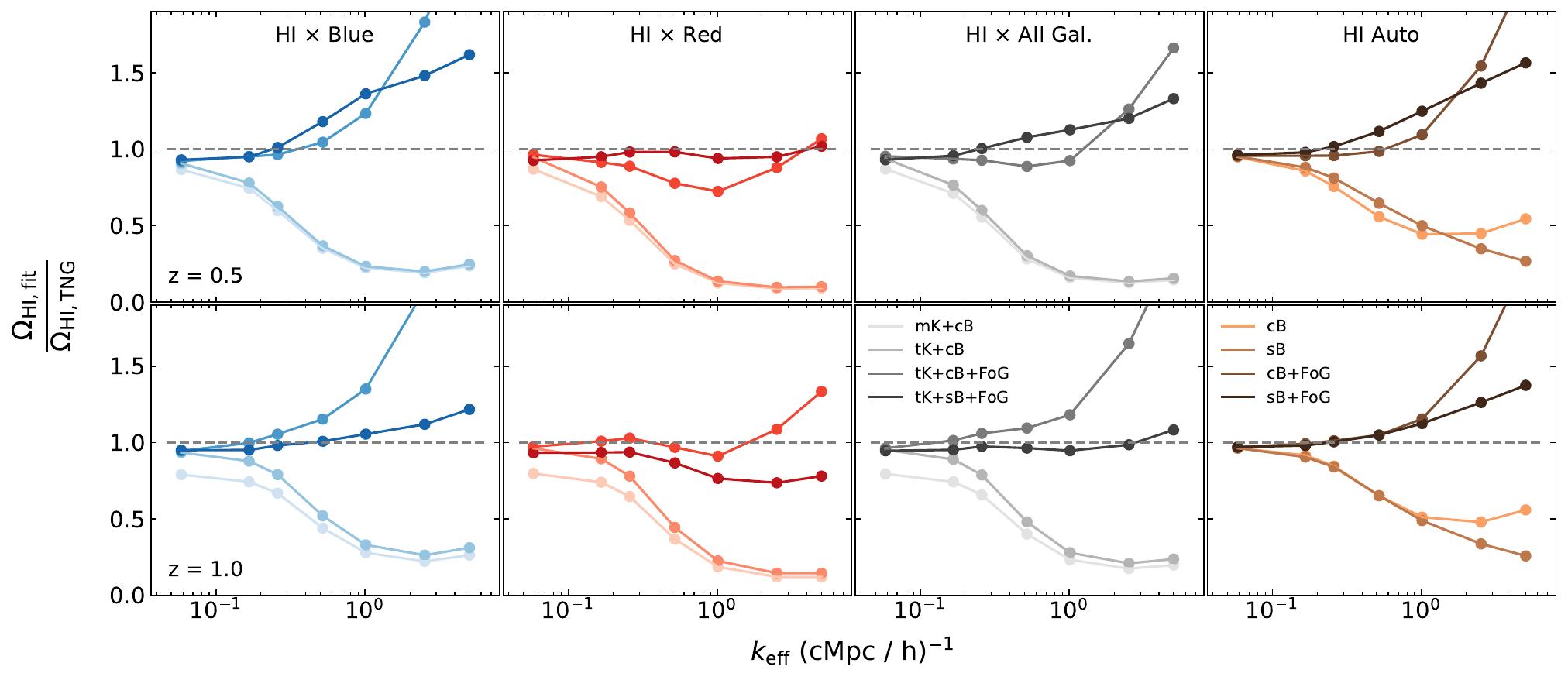}
    \caption{Accuracy of constraints on $\Omega\shi$ at $z = 0.5$ (top) and $z = 1$ (bottom) inferred from fitting various models to \hi power spectra (from left to right: \hib, \hir, \hia, and \hi auto). In practice, models are fit to \hi power spectra over a range of scales and the resulting $\Omega\shi$ value is expressed as a measurement at the average $k$ in that range, \keff. We vary the maximum $k$ of the scales included in the fit to reflect that procedure, expressing the inferred $\Omega\shi$ as a function of \keff. These values are shown as a ratio over $\Omega\shi$ calculated directly from IllustrisTNG, such that values near unity indicate better model performance. Darker colors correspond to more complex models, with the names corresponding to the shorthands and definitions provided in Fig.~\ref{fig:auto_mod}-\ref{fig:xpk_mod}.}
    \label{fig:omega_fits}
\end{figure*}

In summary, Fig.~\ref{fig:xpk_mod} shows that \hi-galaxy cross-power spectra can be interpreted with models that include FoG on scales relevant to recent observations ($k \lesssim 1.5$ \hmpc), given that the acceptable error threshold is $10\%$. However, a model that achieves accuracy even to the $10\%$ level requires exact knowledge of the full scale-dependent bias and pairwise velocity dispersion of the galaxy and \hi populations and a precise measurement of the cross-power spectrum without systematics.

\subsection{Effect on inferred $\Omega\shi$} \label{perf:omega}

% quote from CHIME -- As the constraints with fixed nonlinear parameters do not include the full modeling uncertainties, they show the internal consistency and significance of our measurements but are not good indicators of the plausible range of ΩH I determined from our data. In all cases we are showing constraints derived with the fiducial eBOSS QSO error model. As discussed in Section 8.1, we believe that this model may bias the QSO constraints (particularly the lowest-redshift bin) to higher values of ΩH I.
% makes me feel better about choosing the not fixed values with the huge error bars

In Sections~\ref{perf:auto}-\ref{perf:x}, we showed that models of the form in Equations~\ref{eq:auto_full}-\ref{eq:cross_full} can deviate significantly from the actual \hi auto and cross-power spectrum. Here, we seek to establish how these errors propagate to cosmological constraints made employing these models, using recent measurements of the cosmic abundance of \hi ($\Omega\shi$) as an example. As we will show, these analytical errors can even eclipse the stated measurement uncertainties of the constraints, suggesting that model improvements are a priority. First, we describe the procedure used to infer $\Omega\shi$ from \hi auto and cross-power spectra and compare to the directly-calculated $\Omega\shi$ in Fig.~\ref{fig:omega_fits}. Then, we project the differences between the inferred and actual $\Omega\shi$ onto recent constraints from \citetalias{wolz_h_2022}, \citetalias{cunnington_h_2022}, \citetalias{2023_chime_citation}, \citetalias{paul_first_2023_eprint}, and \citetalias{Carucci24} to illuminate the predicted error from determining $\Omega\shi$ from the \hi power spectra in this way. Again, we emphasize that the tests presented in this section represent the best case for these models since our results are not affected by noise or systematics and the biases, correlation coefficients, and pairwise velocity dispersions are all extracted directly from the simulation.

% start with this section
\subsubsection{Accuracy of $\Omega\shi$ constraints from power spectra} \label{omega:err}

Measurements of $\Omega\shi$ probe the state of star-formation and gas in galaxies over cosmic time and inform galaxy formation models \citep{wyithe_method_2008, diemer_atomic_2019, dave_galaxy_2020}. 21cm intensity mapping experiments can determine $\Omega\shi$ via the amplitude of \hi power spectra since they directly measure brightness temperature fluctuations of \hi, which rescale the density fluctuations by the mean \hi brightness temperature \citep{battye_h_2013}:
\begin{equation}
    \Bar{T}\shi (z) = 180 \Omega\shi (z) h \frac{(1 + z)^2}{\sqrt{\Omega_m (1 + z)^3 + \Omega_\Lambda}} \mathrm{mK} \,.
\end{equation}
Since $\Bar{T}\shi \propto \Omega\shi$, the amplitude of the \hi auto and cross-power spectra are proportional to $\Omega\shi^2$ and $\Omega\shi$, respectively. Estimates of $\Omega\shi$ are obtained by fitting a chosen model to the \hi power spectra over a range of scales. The significance and accuracy of the $\Omega\shi$ constraint changes depending on the included scales due to the trade-off between model accuracy and constraining power. We follow the literature by characterizing the included scales with their average, \keff. 

\begin{deluxetable*}{ccccccccc} \label{tab:ohi}
    \tablecaption{$\Omega\shi$ Constraints from Observations}

    \tablehead{
        \colhead{Paper}                 &
        \colhead{Galaxy Survey}         &          
        \colhead{$z_{\mathrm{eff}}$}    &
        \colhead{\keff}                 &
        \colhead{Model}                 &
        \colhead{$10^3\mathcal{A}\shi$} &
        \colhead{$b\shi$}               &
        \colhead{$r_{\mathrm{HI-Gal}}$} &
        \colhead{$10^3\Omega\shi$} 
    }
    \startdata
        \citetalias{wolz_h_2022} & WiggleZ & 0.78 & 0.24 & \xmA & $0.55\pm0.11$ & 0.8 & 0.9 & $0.94\pm0.26$ \\
        & ELG & 0.78 &  0.24 & \xmA& $0.70\pm0.12$ & 0.8 & 0.7 & $0.96\pm0.27$\\
        & LRG & 0.78 &  0.24 & \xmA &$0.45\pm 0.10$ & 0.8 & 0.6 & $0.96\pm0.29$\\
        \citetalias{cunnington_h_2022} & WiggleZ & 0.43  & 0.13 & \xmA& $0.86\pm0.22$ & $1.13\pm0.1$ & $0.9\pm0.1$ & $0.83\pm0.26$\\
        \citetalias{Carucci24} & WiggleZ & 0.43  & 0.13 & \xmA& $0.93\pm0.17$ & $1.13\pm0.1$ & $0.9\pm0.1$ & $0.91\pm0.21$ \\
        \citetalias{2023_chime_citation} & ELG & 0.96 & 1.0 & \xmC & $4.75^{+9.04}_{-3.80}$ & $1.3\pm0.12$ & 1 & $3.80^{+6.30}_{-1.70}$\\
        & LRG & 0.84 & 1.0 & \xmC & $1.06^{+3.60}_{-0.97}$ & $1.3\pm0.12$ & 1 & $1.05^{+1.83}_{-0.78}$\\
        \citetalias{paul_first_2023_eprint} & -- & 0.44 & 1.0 & \amD & -- & -- & -- & $0.56_{-0.18}^{+0.23}$\\
         & -- & 0.32 & 1.0 & \amD & -- & -- & -- & $0.43_{-0.14}^{+0.18}$
    \enddata
    \tablecomments{The $\Omega\shi$ constraints from \hi-galaxy cross-power spectra reported in \citetalias{wolz_h_2022}, \citetalias{cunnington_h_2022}, \citetalias{2023_chime_citation}, and \citetalias{Carucci24}, and \hi auto power spectra from \citetalias{paul_first_2023_eprint}. The galaxy survey indicates the cross-correlated galaxy population: WiggleZ \citep{blake_wigglez_2011} and ELG \citep{raichoor_preliminary_2020} and LRG \citep{ross_completed_2020} samples from the eBOSS survey. $z_{\mathrm{eff}}$ and \keff represent the effective redshift and wavenumber (in \hmpc) the measurement takes place. The models listed are the closest analog to the model adopted in each work. $10^3\mathcal{A}\shi$ is the quantity directly measured by \citetalias{wolz_h_2022}, \citetalias{cunnington_h_2022}, and \citetalias{Carucci24}. \citetalias{2023_chime_citation} reported $\mathcal{A}\shi = 10^3\Omega\shi (b\shi + \langle f \mu^2 \rangle)$, which we adjust to match Equation~\ref{eq:amplitude}. These works then assume \hi bias ($b\shi$) and \hi-galaxy correlation coefficient ($r_{\mathrm{HI-Gal}}$) values to roughly estimate $10^3\Omega\shi$ from $\mathcal{A}\shi$, sometimes incorporating uncertainties from the assumed values into their $\Omega\shi$ constraints. \citetalias{Carucci24} is an extension of results from \citetalias{cunnington_h_2022}, so we employ the same assumed values although \citetalias{Carucci24} does not directly make such assumptions. Conversely, \citetalias{paul_first_2023_eprint} fit several parameters to their measured \hi auto power spectra and thus do not assume any values, but they do report $\Omega\shi$ values with and without priors in their fits. We adopt their $\Omega\shi$ values from those with priors since they agree more closely with other constraints, but we caution that the uncertainties in $\Omega\shi$ are likely underestimated.}
    % CHIME X LRG actual reported value - $1.51^{+3.60}_{-0.97}$, CHime x ELG $6.76^{+9.04}_{-3.79}$ this is reported with Kaiser term, so needed to remove it manually. (A = OHI(bHI + <f\mu^2>), subtracted out the <> term to just get A in terms of OHI and bHI.
\end{deluxetable*}

We replicate the above procedure by scaling the \hi auto and cross-power spectra by $\Bar{T}\shi (z)$ calculated from $\Omega_{\rm{HI, TNG}}$, determined by simply summing all the \hi in the box. After rescaling, we then fit each of the models studied to the power spectra in Section~\ref{sec:models}. In mathematical terms, we are trying to find a $\Bar{T}_{\mathrm{HI, fit}}$ such that
\begin{align} \label{eq:amp_fit}
    \Bar{T}_{\mathrm{HI, TNG}}^2 P^{\mathrm{TNG}}_{\mathrm{HI}} (k) = \Bar{T}_{\mathrm{HI, fit}}^2 P^{\mathrm{Model}}_{\mathrm{HI}} (k) \,,\\ 
    \Bar{T}_{\mathrm{HI, TNG}} P^{\mathrm{TNG}}_{\mathrm{HI} \times \mathrm{Gal}} (k) = \Bar{T}_{\mathrm{HI, fit}} P^{\mathrm{Model}}_{\mathrm{HI} \times \mathrm{Gal}} (k) \,,
\end{align}
for the auto and cross-power spectra, respectively. We vary the minimum scale included in the fit to examine the small-scale trade-off between the larger errors and smaller uncertainties, expressing the inferred $\Omega\shi$ values as a function of \keff. We follow previous literature to estimate the power spectra uncertainties used in the fits \citep{furlanetto_cross-correlation_2007, lidz_probing_2009, park_cross-power_2014}, removing any instrumental contributions. We compare the value determined in the fit, $\Omega_{\rm{HI, fit}}$, to $\Omega_{\mathrm{HI, TNG}}$, providing the results for each model of the \hi auto and cross-power spectrum in Fig.~\ref{fig:omega_fits}. We exclude error bars from uncertainties in the fitted values in Fig~.\ref{fig:omega_fits} for clarity (error bars included in online figures), but we note that all uncertainties are negligible except for the value at the smallest \keff.   
 
%TODO should I specify in the bias/cc sections that I use the large-scale limits for the paper, not those in the table

Ideally, the $\Omega_{\mathrm{HI, fit}}/\Omega_{\mathrm{HI, TNG}}$ ratios of all models should approach unity on large scales, given that the largest scales probed by TNG300 should be linear \citep{smith_stable_2003}. The models in Fig.~\ref{fig:omega_fits} do converge on large scales, with one exception: \xmA, which at $\keff \approx 0.05$ \hmpc only reaches a ratio of about 0.8 and 0.9 for $z = 1$ and $z = 0.5$, respectively. These errors would significantly skew the recent constraints of $\Omega\shi$ that resorted to this model (see Section~\ref{omega:revisions}). Interestingly, the remaining models also do not converge quite to unity, instead converging to $\approx 0.97$ and $\approx 0.94$ for the auto and cross-power spectra, respectively. These large-scale differences may imply that there is an absent large-scale effect in the models, although we reserve further analysis to future work with a larger-volume simulation.

This slight vertical offset between unity and the $\Omega_{\mathrm{HI, fit}}/\Omega_{\mathrm{HI, TNG}}$ ratios serve as the largest difference between Fig.~\ref{fig:omega_fits} and the ratios from Fig.~\ref{fig:auto_mod} and \ref{fig:xpk_mod}. Otherwise, the $\Omega_{\mathrm{HI, fit}}/\Omega_{\mathrm{HI, TNG}}$ errors are effectively the reciprocal of $P_{\mathrm{Model}} / P_{\mathrm{TNG}}$, as expected from Equation~\ref{eq:amp_fit}. Beyond this first-order effect, $\Omega_{\mathrm{HI, fit}}/\Omega_{\mathrm{HI, TNG}}$ tends to deviate from its large-scale value at smaller \keff than its $P_{\mathrm{Model}} / P_{\mathrm{TNG}}$ counterpart evaluated at a similar $k$. This trend suggests that the larger errors at $k > \keff$ have a stronger impact than the smaller ones at $k < \keff$ in the determination of $\Omega_{\mathrm{HI, fit}}$. 

\begin{figure*}
    \centering
    \includegraphics[width=\linewidth]{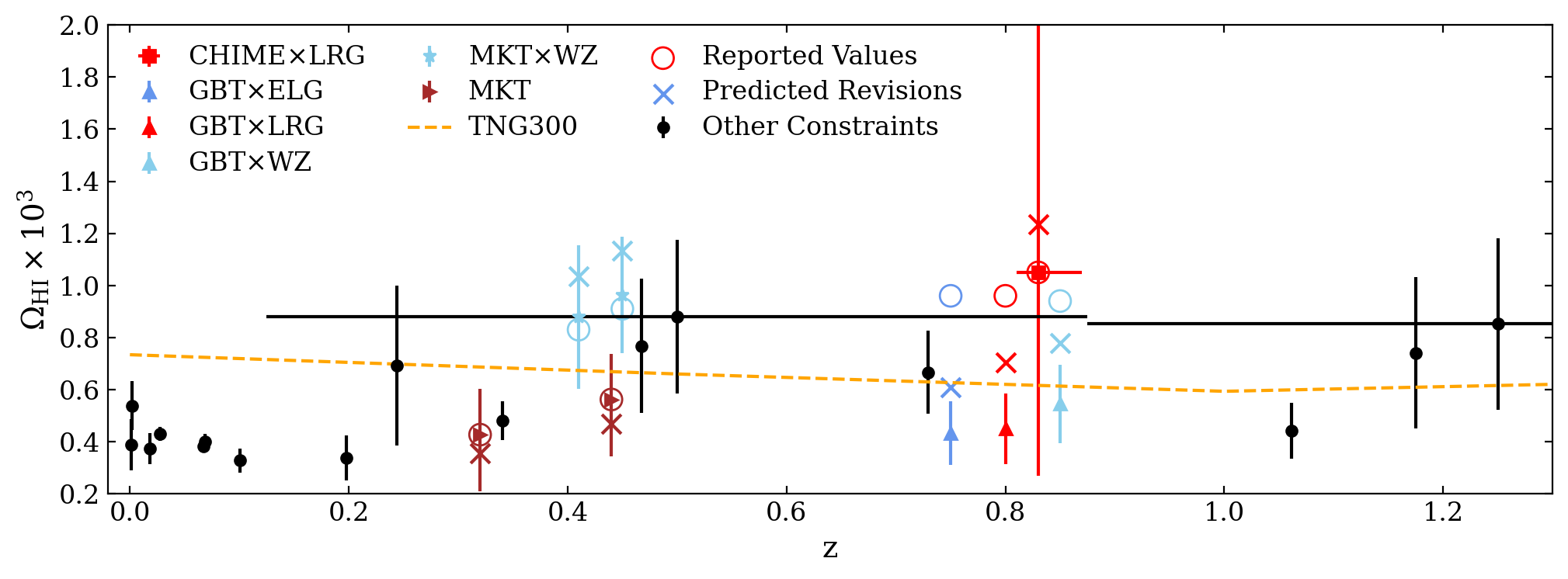}
    \caption{Comparisons of reported $\Omega\shi$ constraints from \citetalias{wolz_h_2022} (vertical triangles, horizontally offset for illustrative purposes), \citetalias{2023_chime_citation} (square), \citetalias{paul_first_2023_eprint} (horizontal triangles), and MeerKAT-WiggleZ constraints (\citetalias{cunnington_h_2022} and \citetalias{Carucci24}, stars offset to left and right, respectively) to the $\Omega\shi$ values revised according to the predicted model errors at the nearest $z$ from IllustrisTNG (crosses). The significant differences between the crosses and points demonstrate the substantial analytical error incurred by the models used to interpret these measurements. We emphasize that these represent the minimum possible error inherited from the model, including some instrumental effects for \citetalias{wolz_h_2022}, \citetalias{cunnington_h_2022}, and \citetalias{Carucci24}. There are caveats that impact the interpretation of results from \citetalias{2023_chime_citation} and \citetalias{paul_first_2023_eprint} (see text and Table~\ref{tab:ohi}). Constraints on $\Omega\shi$ from other \hi measurement techniques and $\Omega\shi$ values from IllustrisTNG are shown as black points and a yellow dashed line, respectively. $\Omega\shi$ is extracted from the measured $\mathcal{A}\shi$ (Equation~\ref{eq:amplitude}) using $b\shi$ and $r_{\mathrm{HI-Gal}}$ from IllustrisTNG which can differ from those used in the respective works (Table~\ref{tab:ohi}). We display the originally reported $\Omega\shi$ values as open circles for completeness but note that these are not directly comparable since each adopts different $b\shi$ and $r_{\mathrm{HI-Gal}}$ values.}    
    \label{fig:omhi_revisions}
\end{figure*}

% discuss redshift evolution at all? not sure if there's really anything that interesting
\subsubsection{Predicted impact on recent observations} \label{omega:revisions}

We now contextualize the errors found in Section~\ref{omega:err} by showing how recent $\Omega\shi$ constraints (Table~\ref{tab:ohi}) would change given the predicted errors from IllustrisTNG in Fig.~\ref{fig:omhi_revisions}. However, interpreting the observations is challenging because the \hi bias and \hi-galaxy correlation coefficient are unknown \textit{a priori} and both quantities are also proportional to the power spectrum amplitude (Equations~\ref{eq:auto_full}-\ref{eq:cross_full}). As a result, the auto and cross-power spectra probe the degenerate quantities $\Omega\shi^2 b\shi^2$ and $\Omega\shi b_{\mathrm{Gal}} b\shi r_{\mathrm{HI-Gal}}$, respectively. These degeneracies can be broken in the quadrupole and hexadecapole (Equation~\ref{eq:poles}) in future surveys with improved signal-to-noise, but current observations of the cross-power spectrum resort to constraints of the quantity 
\begin{equation}\label{eq:amplitude}
    \mathcal{A}\shi = \obr \,,
\end{equation}
taking $b_{\mathrm{Gal}}$ from the galaxy survey. $\Omega\shi$ can be roughly estimated from $\mathcal{A}\shi$ by assuming values of $b\shi$ and $r_{\mathrm{HI-Gal}}$ from theoretical predictions, which can vary amongst the different works (values given in Table~\ref{tab:ohi}). We replace the $b\shi$ and $r_{\mathrm{HI-Gal}}$ estimates from observations with those from TNG300 to enable consistent comparisons in Fig.~\ref{fig:omhi_revisions}, but we still present the original $\Omega\shi$ values for completeness (open circles). We then use the estimate of the corresponding model error from IllustrisTNG at the nearest $z$ ($\Omega_{\mathrm{HI, fit}}/\Omega_{\mathrm{HI, TNG}}$ values from Fig.~\ref{fig:omega_fits}) to predict what the ``actual'' $\Omega\shi$ values would be (crosses) if the measurements were interpreted with a perfect model. We note that $z_{\rm{sim}} > z_{\rm{eff}}$ for each observation, which usually underestimates model errors since they typically grow with redshift across $0 \leq z \leq 1$.

However, the \hi distribution in IllustrisTNG would not be perfectly measured by 21cm intensity mapping experiments due to re-gridding \citep{cunnington_accurate_2024} and the beams of the instruments themselves. In previous sections, we remained agnostic to instrument-specific effects in order to keep our results general and to focus on errors originating from models. In Fig.~\ref{fig:omhi_revisions}, however, we examine the impact of model errors on the $\Omega\shi$ constraints from specific observations. To imitate how each instrument would observe IllustrisTNG's \hi distribution, we apply a Gaussian filter to the simulated redshift-space \hi density grid, following \citet{wolz_determining_2017}:
\begin{align}
    G(x,y) &= \frac{1}{2\pi R_{\text{beam}}^2} \exp\left(-\frac{x^2 + y^2}{2 R_{\text{beam}}^2} \right), \\
    R_{\text{beam}} &= \frac{d_c \cdot c}{N_{\rm{grid}} \nu D_{\rm{dish}} 2 \sqrt{2 \ln 2}}.
\end{align}
Here, $d_c$, $c$, $N_{\rm{grid}}$, and $\nu$ are the co-moving distance, speed of light, number of grid points along one axis, and redshifted 21cm line, respectively. $D_{\rm{dish}}$ represents the diameter of the dish for each radio telescope --- we use 100m for Greenbank Telescope \citep[GBT, ][]{masui_measurement_2013, switzer_determination_2013} and 13.5m for MeerKAT \citep[MKT, ][]{wang_h_2021}. We assume a constant beam size along the line of sight, as the box is at one redshift, and apply the smoothing on each two-dimensional slice perpendicular to the line-of-sight ($z$) independently.

%To more accurately represent how model errors would affect their constraints, we apply a simple beam to the \hi distributions to imitate how they would appear to the Greenbank Telescope \citep[GBT,][]{masui_measurement_2013, switzer_determination_2013} and MeerKAT \citep[MKT,][]{wang_h_2021}. This 

We also adjust our model to accommodate the beam, following \citet{ando_redshift_2019}:
\begin{equation}
    P_{\rm{HI} \times Gal}^{\rm{obs}} (k, \mu) = W_{\rm{beam}} (k, \mu, \sigma_{\rm{sm}}) P_{\rm{HI} \times Gal}(k, \mu) \,,
\end{equation}
where $\sigma_{\rm{sm}}$ represents the angular resolution of each instrument at the redshift of interest. Overall, including the beam reduces the error estimates by a factor of $\gtrsim$0.95 for \citetalias{cunnington_h_2022}, \citetalias{wolz_h_2022}, and \citetalias{Carucci24}. We do not employ this procedure for \citetalias{2023_chime_citation} or \citetalias{paul_first_2023_eprint}, as they adopt more complicated pipelines to interpret their data. We instead directly apply the error estimates from Fig.~\ref{fig:omega_fits}; the corresponding results shown in Fig.~\ref{fig:omhi_revisions} should be understood as the lower limit on the magnitude of errors originating from the model\footnote{\citetalias{paul_first_2023_eprint} adopted a halo model in their \hi auto power spectrum model, which requires additional assumptions on how \hi occupies halos. If these assumptions are valid, then \amD is an appropriate equivalent (Table~\ref{tab:ohi}).}, excluding any instrumental effects.
% Consequently, reported $\Omega\shi$ from observations (which we show with open circles in Fig.~\ref{fig:omhi_revisions}) can differ because of the assumed $b\shi$ and $r_{\mathrm{HI-Gal}}$ values. We substitute the IllustrisTNG values for these quantities to remove these differences We remove the this effect by substituting the $b\shi$ and $r_{\mathrm{HI-Gal}}$ values from IllustrisTNG, and display the resulting $\Omega\shi$ values in Fig.~\ref{fig:omhi_revisions}. For completeness, we also show the original $\Omega\shi$ values reported in each work, but  %We investigate the impact of these assumptions by making two comparisons in the top panel of Fig.~\ref{fig:omhi_revisions}. First, we study how the model errors impact constraints --- this should manifest in both $\mathcal{A}\shi$ and $\Omega\shi$ since model errors shift the interpreted amplitude of the \hi auto and cross-power spectra. Second, we replace the $b\shi$ and $r_{\mathrm{HI-Gal}}$ values with those from IllustrisTNG to facilitate comparisons to $\Omega\shi$ constraints from other techniques in the bottom panel.

Fig.~\ref{fig:omhi_revisions} shows that the employed models significantly bias the inferred $\Omega\shi$ even with respect to the stated uncertainties for each measurement. The magnitude of the error can vary with the model and redshift. For example, the \xmA model incurs larger errors for \citetalias{wolz_h_2022} than \citetalias{cunnington_h_2022} and \citetalias{Carucci24} because the matter Kaiser term assumption is safer for late-time biases that approach unity (Table~\ref{tab:bias}). Conversely, \citetalias{2023_chime_citation} adopts a more complex model that resembles \xmC, which somewhat mitigates the analytical error at the expense of larger uncertainties. 

Fig.~\ref{fig:omhi_revisions} contains $\Omega\shi$ constraints from three techniques other than \hi intensity mapping \citep[black,][]{zwaan_hipass_2005, rao_damped_2006, lah_hi_2007, martin_arecibo_2010, braun_cosmological_2012, rhee_neutral_2013, hoppmann_blind_2015, rao_statistical_2017, jones_alfalfa_2018, bera_atomic_2019, hu_accurate_2019, chowdhury_h_2020}. At low redshifts $(z \approx 0)$, the \hi content of galaxies is measured directly in blind surveys, obtaining $\Omega\shi$ by integrating the observed \hi mass function. Detections of individual galaxies in a blind survey are difficult at $0.1 \lesssim z \lesssim 1$, so the \hi of optically-selected galaxies is measured via stacking, with $\Omega\shi$ measurements requiring corrections for incompleteness. Finally, at high redshifts ($z \gtrsim 1$), damped Lyman-$\alpha$ absorption yield column densities that can be integrated to produce $\Omega\shi$.

Overall, the revisions in Fig.~\ref{fig:omhi_revisions} improve the agreement between the $\Omega\shi$ constraints from intensity mapping and other techniques. This is particularly true for the constraints from \citetalias{wolz_h_2022}, while revised $\Omega\shi$ values from \citetalias{cunnington_h_2022}, \citetalias{2023_chime_citation}, and \citetalias{Carucci24} are still consistent with the other constraints. The $\Omega\shi$ constraint from CHIME $\times$ ELG is revised downward, although it is not shown in Fig.~\ref{fig:omhi_revisions} due to its unusually large value --- \citetalias{2023_chime_citation} attribute this to a chance fluctuation and the prior volume effect (see their section 6.3).

Even when considering the ideal, best-case limit of these models, the revisions in Fig.~\ref{fig:omhi_revisions} demonstrate that current constraints can be limited by systematics in the analysis rather than the signal-to-noise of the measurements themselves. Clearly, more sophisticated models are needed (more in Section~\ref{disc:fog}) in order to properly realize the capabilities of future 21cm intensity mapping experiments with improved signal-to-noise.

\section{Discussion} \label{sec:disc}

\subsection{What is the impact of simulation technique on the hydrogen distribution?} \label{disc:sims}

\begin{deluxetable*}{ccccccc}
\tablecaption{\hi Bias Values from Various Simulations}
\label{tab:bias_compare}
\tablewidth{0pt}
\tablehead{
\colhead{Source} & \colhead{Methodology} & \colhead{Post-Processing} & \colhead{$m_{\mathrm{dm}} (M_\odot)$} & \colhead{$k_{\mathrm{eff}}$ (\hmpc)} & \colhead{Other Cuts ($M_\odot$)} & \colhead{$b\shi (z = 1)$}
}
\startdata
This work & \hydro & - & 5.9 $\times 10^7$ & 0.07 & - & 1.28 \\
\citet{ando_redshift_2019} & \hydro & - & 2.9 $\times 10^7$ & 0.26 & - & 1.22\\
\citetalias{villaescusa-navarro_ingredients_2018} & \hydro & - & 7.5 $\times 10^6$ & 0.15 & - & 1.49 \\
\hline
\citet{wolz_intensity_2016} & \nbody & SAM & $1.2 \times 10^9$ & $0.1$ & $M_h \geq 10^{10}$ & $\sim 1.8$ \\
\citet{spinelli_atomic_2020} & \nbody & SAM & $1.2 \times 10^9$ & $0.015$ & $M_h \gtrsim 10^{10}, M_\star \gtrsim 10^8$ & 1.22 \\
\citet{wang_breakdown_2021} & \nbody & SAM & $5 \times 10^8$ & 0.05 & $M_h \geq 1.5 \times 10^{10}$ & 1.26 \\
\citet{guo_neutraluniversemachine_2023} & \nbody & SAM & $2 \times 10^8$ & 0.05 & $M_{\mathrm{HI}} \geq 10^{6.9}$ & 1.10 \\
\citet{spinelli_atomic_2020} & \nbody & SAM & $10^7$ & $0.1$ & $M_h \gtrsim 10^{8}, M_\star \gtrsim 10^6$ & 1.31 \\
\hline
\citet{wang_breakdown_2021} & \nbody & HIHM & $5 \times 10^8$ & 0.05 & $M_h \geq 1.5 \times 10^{10}$ & 1.48 \\
\citet{sarkar_modelling_2016} & \nbody & HIHM & $10^8$ & 0.065 & - & $\sim 0.9$ \\
\citetalias{villaescusa-navarro_ingredients_2018} & \nbody & HIHM & 7.5 $\times 10^6$ & 0.15 & - & $\sim 1.2$ \\
\hline
\citet{penin_scale-dependent_2018} & \anal & - & - & 0.01 & - & $\sim$0.82 \\
\citet{umeh_imprint_2017} & \anal & - & - & 0.01 & - & $\sim$0.86 \\
\enddata
\tablecomments{Summary of the $z \approx 1$ \hi bias values found in various computational and theoretical works. The values are grouped by methodology (see text for explanation) and ordered from top to bottom by increasing mass resolution (fourth column). Some works make additional mass cuts to avoid including poorly resolved structures (sixth column). $k_{\rm{eff}}$ (fifth column) represents the approximate scale at which the bias was measured, although the impact of this should be small since the $z = 1$ \hi bias is nearly constant to very small scales in all works. Bias values averaged over different models or estimated are denoted with $\sim$. We aim to establish rough trends and thus do not report statistical uncertainties arising from cosmic variance for simplicity.}
\end{deluxetable*}

In Section~\ref{sec:perf}, we found that IllustrisTNG predicts that general large-scale models (Equations~\ref{eq:auto_full}-\ref{eq:cross_full}) introduce significant errors in cosmological constraints. However, simulated \hi distributions can be sensitive to the included galaxy formation physics \citep{park_cross-power_2014, Li24_eprint, Marasco25}. This raises a key question: does IllustrisTNG's galaxy formation model inherently yield complicated \hi distributions? We explore this question by comparing results from some selected computational and theoretical works in Table~\ref{tab:bias_compare}, broadly characterized by their $z = 1$ \hi bias.

We examine results from three methodologies: \hydro, \nbody, and \anal. The first two, \hydro and \nbody, describe simulations that differ in their handling of baryons. Hydrodynamical simulations like IllustrisTNG include baryonic physics on-the-fly, whereas N-body simulations defer baryon modeling to post-processing, typically through a \hi-halo mass relation (HIHM) or semi-analytic model (SAM). In the remaining methodology, \anal, perturbation theory \citep{bernardeau_large-scale_2002} and a halo model are employed to study \hi distributions without a simulation.

%This technique is computationally simple, but neglects the dependence of \hi mass on secondary halo properties such as age \citep{gao_age_2005} and removes the contribution of clustering within halos.  
%better account for the impact of assembly history, but require the tuning of many input parameters.
%, neglecting second-order nonlinear contributions to \hi clustering and the dependence of a halo's \hi mass on secondary halo properties for the sake of speed and simplicity

\hi bias values in the first methodology, \hydro, appear to agree across different simulations at $z = 1$ \citep[also see figure 2 from][]{ando_redshift_2019}, although they show notable dependence on resolution. We note that our model tests are robust against these resolution effects since they are dependent on the \hi bias shape, rather than its precise value (Appendix~\ref{app:resolution}). We attribute the \hydro resolution dependence to differences in how \hi occupies the largest halos. \citetalias{villaescusa-navarro_ingredients_2018} found that massive halos in coarser simulations tended to contain less \hi than finer ones, in turn suppressing \hi clustering. This agrees with other works that studied the resolution dependence of star-formation in IllustrisTNG. For example, \citet{donnari_quenched_2021-1} found that the quenching of the massive satellites ($M_\star \geq 10^{10} M_\odot$) in these massive host halos is enhanced in the coarser IllustrisTNG simulations, in turn suggesting reduced \hi abundance.  %One potential explanation for this is ram-pressure stripping, which preferentially occurs in massive halos, is stronger in coarser resolutions \citep{gunn_infall_1972,hester_ram_2006,hes_gas_2012, steinhauser_simulations_2016, donnari_quenched_2021-1}. Alternatively, \citet{dave_neutral_2013} speculated that high-resolution simulations have enhanced star-formation rates that may also reduce \hi abundance.

SAMs exhibit a similar resolution trend in their \hi clustering, albeit for different reasons. Instead of a reduction of \hi occupation in large halos, SAMs of coarser-resolution simulations possess enhanced \hi occupation in small halos \citep[see figure 6 in][]{spinelli_atomic_2020}. In both SAMs and \hydro, \hi occupies smaller halos on average, suppressing its clustering. The \hi bias value also depends on the particular SAM implementation, as demonstrated by  \citet{wolz_intensity_2016} and \citet{guo_neutraluniversemachine_2023}. Identifying the particular SAM components that cause these differences is left to future work.

HIHMs, conversely, have a non-trivial dependence on resolution and the employed HIHM. The resolution dependence is clearest when comparing \citetalias{villaescusa-navarro_ingredients_2018} and \citet{wang_breakdown_2021}, which adopt the same HIHM but still find \hi biases that differ by $\sim 0.3$. \citet{wang_breakdown_2021} reasoned that the discrepancy arises because halos below their resolution threshold contribute non-negligibly to the overall \hi distribution. However, even at similar resolutions, \citet{sarkar_modelling_2016} and \citet{wang_breakdown_2021} find biases that differ by $\sim 0.5$, implying that the \hi bias is also dependent on the adopted HIHM. 
%. This disagreement suggests that the \hi bias is sensitive to the applied HIHM as well, conflicting with conclusions from \citet{penin_scale-dependent_2018} who found differences of only $\sim 0.05$ between the \hi biases from various HIHM models. A potential solution to the disagreement between \anal and HIHMs is that the absent small halos in \citet{sarkar_modelling_2016} and \citet{wang_breakdown_2021} cause an increased sensitivity in the \hi bias to small differences in HIHMs.

Comparisons among the methodologies indicate that \hi bias values are sensitive to secondary halo properties, such as halo age or assembly history \citep{gao_age_2005}. Methodologies like \hydro and SAMs that natively incorporate these assembly effects generally converge within $1.2 < b\shi < 1.5$, particularly at comparable resolutions \citep{ando_redshift_2019}. \citet{marasco_environmental_2016} found that SAMs have stronger \hi occupation in dense environments as compared to \hydro, which may influence the remaining differences between \hydro and SAMs. Conversely, methodologies that neglect assembly effects are either substantial outliers from \hydro and SAMs, like \anal, or vary dramatically between implementations, like HIHM. However, in the case of HIHM, \citetalias{villaescusa-navarro_ingredients_2018} showed that its overall normalization and the fraction of \hi outside of halos can also contribute to this disparity in behavior. 
%The final methodology, \anal, contains works that report values of $b\shi \sim 0.84$, which stand as outliers from the \hydro and SAM results that generally lie within $1.2 < b\shi < 1.5$. This suggests that the value of the \hi bias is sensitive to secondary halo properties, such as halo age or assembly history \citep{gao_age_2005}. This conclusion is further supported in direct comparisons of HIHMs and \hydro results --- both \citet{ando_redshift_2019} and \citetalias{villaescusa-navarro_ingredients_2018} found significant differences between their \hydro and HIHM biases.

In summary, Table~\ref{tab:bias_compare} shows that the predicted amplitude of the $z = 1$ \hi bias is dependent on the methodology and resolution. Throughout the section, we have provided some speculative reasons for these trends. Conversely, the shape of the \hi bias is qualitatively similar throughout the various works, which indicates that our results are robust against these effects. A deeper study that quantitatively establishes these methodology and resolution relationships is needed to develop and test more sophisticated models on realistic \hi distributions at $z \leq 1$.

\subsection{Why is the FoG effect such a large source of error?} \label{disc:fog}

\begin{figure}
    \centering
    \includegraphics[width=.65\linewidth]{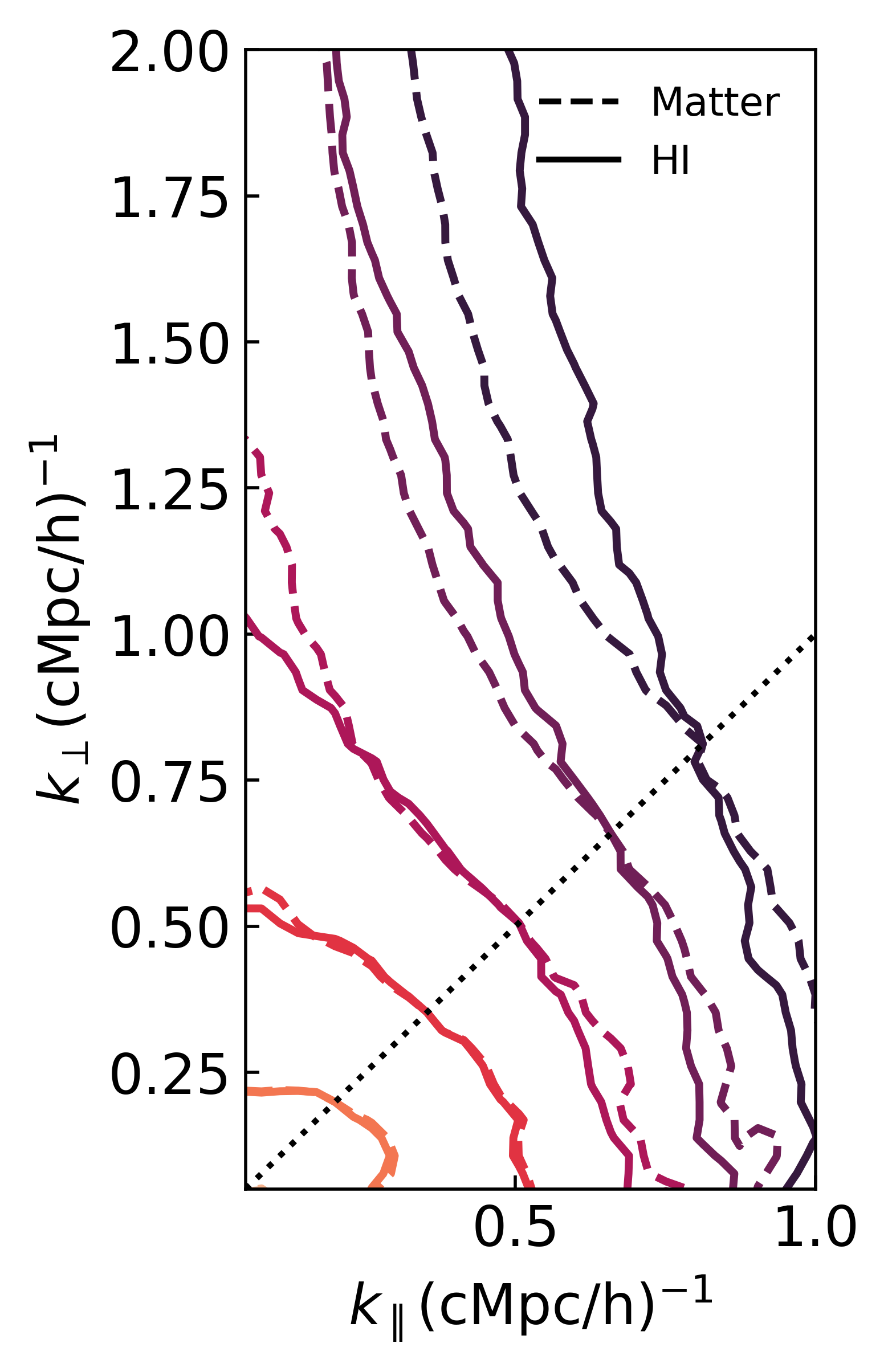}
    \caption{Comparison of the strength of FoG suppression for \hi ($\sigma\shi$) and matter ($\sigma_{\mathrm{m}}$) at different scales, with lighter colors corresponding to larger scales. The isopower contour lines delineate, at some $k$, $P_{\mathrm{HI}}(k_\parallel = k, k_\perp = k)$ and $P_{\mathrm{m}}(k_\parallel = k, k_\perp = k)$ such that they intersect at points along the dotted line. Defining contours in this way enables easier comparisons of the contour shapes, although the matter and \hi contours represent different power values and spacings. We smooth the contours with a Gaussian filter ($\sigma = 0.8$) to reduce noise, which does not impact our conclusions. \hi contour lines are more vertically stretched than matter at $k \lesssim 0.75$ \hmpc, implying stronger FoG suppression. Conversely, contour lines starting at $k \gtrsim 0.75$ \hmpc exhibit the opposite trend: matter is stretched more than \hi.}
    \label{fig:fog_dif}
\end{figure}

Throughout Sections~\ref{sec:ing}-\ref{sec:perf}, we found considerable evidence that an FoG term like Equation~\ref{eq:fog} is a poor fit to the auto power spectra, resulting in our so-called ``RSD errors''. In this section, we further demonstrate with Fig.~\ref{fig:fog_dif} that models like Equations~\ref{eq:auto_full}-\ref{eq:cross_full} may insufficiently describe \hi RSDs, speculate on potential reasons for these issues, and conclude by proposing adjustments to these models.

We compare in Fig.~\ref{fig:fog_dif} the relative FoG damping strength in \hi and matter, finding that \hi contains stronger damping than matter on large scales and vice versa on small scales. However, Equations~\ref{eq:auto_full}-\ref{eq:cross_full} contain a single damping term whose strength is controlled by a scale-independent quantity, the pairwise velocity dispersion (PVD). Since matter possesses the larger PVD, we would expect that matter experiences stronger damping at \textit{all} scales. Fig.~\ref{fig:fog_dif} conflicts with this picture, suggesting instead that multiple RSD damping contributions are folded into the ``FoG'' term.

One known auxiliary damping source is the neglected nonlinear Kaiser terms in Equations~\ref{eq:auto_full}-\ref{eq:cross_full}, which are typically included as corrections to the matter power spectrum \citep{scoccimarro_redshift-space_2004, taruya_baryon_2010}. \citet{ando_redshift_2019} demonstrated that strictly linear models have reduced PVDs relative to models with these terms (see their figure 5). However, these models necessarily assume that nonlinearities manifest the same in \hi and matter (i.e., no \hi velocity bias), and furthermore including these nonlinearities can actually worsen model performance at $0.5 \lesssim k \lesssim 1$ \hmpc. Given these complications, it is uncertain if the nonlinear contributions to the Kaiser term alone can capture the trends seen in Fig.~\ref{fig:fog_dif}.

Differences in the small- and large-scale behavior of the \hi PVD may also give rise to the behavior in Fig.~\ref{fig:fog_dif}. The \hi PVD can be understood in a halo framework as contributions of three terms, each dominating at different scale regimes: inter-halo (large scales), intra-halo (intermediate), and intra-galactic (small) velocity dispersions \citep{slosar_pairwise_2006, sarkar_redshift-space_2019}. The last contribution is unique to \hi intensity maps as compared to point-source tracers like galaxies, which would not have intra-galactic dispersions \citep[\citetalias{paul_first_2023_eprint}; ][]{li_modelling_2024}. \citet{osinga_atomic_2024} showed that the \hi power spectrum with and without these intra-galactic dispersions diverge at $k \sim 0.5$ \hmpc in IllustrisTNG, with the latter asymptoting to a constant shot noise floor. This comparison demonstrates that intra-galactic dispersions become important at $k \gtrsim 0.5$ \hmpc and the real-space shot noise term is damped \textit{only} by intra-galactic dispersions\footnote{Strictly speaking, shot noise only manifests on scales where random noise emerges and thus cannot be ``damped'' by RSDs. A more precise interpretation is that \hi fills a larger fraction of intra-halo spaces in redshift space, effectively lowering the shot noise floor compared to real space. Consequently, the actual redshift-space shot noise floor should resurface at smaller scales.}. Given this conclusion, we propose that \hi power spectra be modeled phenomenologically with two separate FoG damping terms:
\begin{multline}
    P^S\shi (k, \mu) =  K\shi P\shi^R(k) D_{\mathrm{FoG}}^2 (k, \mu, \sigma\shi) + \\ P^{\mathrm{SN}}\shi D_{\mathrm{FoG}}^2 (k, \mu, \sigma_{\mathrm{int}})  \,. 
    \label{eq:hi_sn}
\end{multline}
Here, $P^{\mathrm{SN}}$ symbolizes the shot noise term from the real-space \hi auto power spectrum. $\sigma\shi$ represents the \hi PVD, which would include all three inter-halo, intra-halo, and intra-galactic components, whereas $\sigma_{\mathrm{int}}$ only includes the intra-galactic dispersions. The two distinct damping terms would reproduce the behavior in Fig.~\ref{fig:fog_dif} if $\sigma_{\mathrm{int}} < \sigma_{\mathrm{m}} < \sigma\shi$, assuming that shot noise is not significant to the matter power spectrum. A thorough understanding of the distribution of the galactic \hi emission profiles can improve the modeling of $\sigma_{\mathrm{int}}$ \citep{li_modelling_2024}.

Disentangling the contributions from the nonlinear Kaiser and scale-dependent PVD to the \hi FoG damping and the testing of Equation~\ref{eq:hi_sn} is left for future work. However, the analysis of Fig.~\ref{fig:fog_dif} illustrates the need for an improved understanding of how RSDs manifest in \hi distributions.

\section{Conclusion} \label{sec:conc}

We have evaluated the performance of models of the large-scale \hi distribution against the hydrodynamical simulation IllustrisTNG. Specifically, we compare predictions for $z \leq 1$ \hi auto and cross-power spectra with blue, red, and the whole galaxy population. This analysis is completed in two phases. First, we extract the model ingredients from the simulation and assessed the validity of common assumptions regarding the functional form of each term. From those ingredients, we construct model power spectra and compare them to the simulated power spectra. Based on these deviations, we predict the impact of these models on inferred constraints on the cosmic abundance of \hi ($\Omega\shi$). Generally, we find that models, even in the best-case scenario, struggle to accurately capture \hi auto and cross-power spectra on scales relevant to observations, such that the model introduces errors that are comparable to current measurement uncertainties. Specifically, we conclude the following:

\begin{enumerate}

    \item The error from assuming a constant bias reaches $\gtrsim 10\%$ at $k \sim 0.2-0.3$ \hmpc for \hi and all galaxy populations. The exceptions are the $z = 1$ \hi bias and $z = 0$ blue galaxy bias, which remain constant at $<10\%$ level until $k \sim 1$ \hmpc because halo occupation effects and nonlinear clustering coincidentally cancel.

    \item The error from assuming a constant \hi-blue galaxy correlation coefficient reaches $10\%$ at $k \sim 0.3$ \hmpc. Conversely, \hi-red deviates from its large-scale value by $>10\%$ on all scales probed by IllustrisTNG ($k \gtrsim 0.05$ \hmpc).

    \item Models with a linear Kaiser term and a single FoG damping term overestimate the corresponding redshift-space power spectra by $\sim 10\%$ at $k \sim 0.15$ \hmpc for all power spectra studied. We find that the \hi pairwise velocity dispersion has different large- and small-scale behavior and propose an adjustment to how RSDs are modeled (Section~\ref{disc:fog}).

    \item We test two common explicit assumptions made to simplify models of tracer power spectra: neglecting the FoG term and assuming a constant bias. Of the two, neglecting FoG suppression causes the largest error in \hi auto and \hi-galaxy cross-power spectra, reaching $\sim 10\%$ by $k \sim 0.1$ \hmpc in both. Assuming a constant bias causes a $\sim 10\%$ error by $k \sim 0.2$ \hmpc.

    \item Even with perfect knowledge of the tracer bias and RSD parameters, the tested models still deviate from the actual \hi auto and \hi-galaxy cross-power spectra by $\gtrsim 10\%$ at scales relevant for observations.

    \item We estimate the error in recent $\Omega\shi$ constraints from models used to interpret recent \hi-galaxy cross-power spectra constraints. We find that the model errors significantly skew the obtained $\Omega\shi$ values by 15-30\%, larger than some measurements' statistical uncertainties. Accounting for these model errors The revisions improve agreement in $\Omega\shi$ values from cross-power spectra and other techniques.

\end{enumerate}

These results have important ramifications for analytical treatments of \hi auto and cross-power spectra at low redshift. Even in the ideal scenario, where each model ingredient is measured exactly, our tested large-scale \hi models introduce systematic errors comparable to current statistical uncertainties in the constraints of cosmological values. The performance of these models is limited by how they model RSDs, suggesting that future analytical work should focus on improving our understanding of how RSDs manifest in tracer distributions.

However, we only studied the performance of these models using a single suite of simulations. Completing similar studies with other simulations would test how its architecture impacts the resulting tracer distributions and the performance of large-scale \hi models. Furthermore, we only examine how models perform in the monopole --- higher-order cross-correlations between \hi and galaxies can provide even more constraining power, but require even more sophisticated models \citep[e.g.,][]{witzemann_simulated_2019, Chand_24}.

%% Please use the acknowledgment and contribution environments. This will 
%% be anonomyized when the "anonymous" style option is used. 
\begin{acknowledgments}
We would like to thank Steven Cunnington and Laura Wolz for their advice, comments, and helpful resources in replicating their work. We also thank Alkistis Pourtsidou and Yi-chao Li for their useful discussions, and the anonymous referee for the helpful comments. Research was performed in part using the compute resources and assistance of the UW-Madison Center For High Throughput Computing (CHTC) in the Department of Computer Sciences. The CHTC is supported by UW-Madison, the Advanced Computing Initiative, the Wisconsin Alumni Research Foundation, the Wisconsin Institutes for Discovery, and the National Science Foundation. We also acknowledge the University of Maryland supercomputing resources (http://hpcc.umd.edu) made available for conducting the research reported in this paper. This research was supported in part by the National Science Foundation under Grant numbers 2206690 and 2338388. BD furthermore acknowledges support from the Sloan Foundation.
\end{acknowledgments}

\begin{contribution}
%%This section gives authors the space to recognize author contributions. The text inside this environment is NOT counted towards the total word quanta. At a minimum, manuscripts are expected to include this text:

The analysis, writing, and coding was led by CKO. BD wrote the software that created the \hi catalogs for IllustrisTNG. FVN provided code that calculated power spectra. Both BD and FVN also advised the project.

%% But authors are expected to provide more specific details, e.g. 
%%
%%SC was responsible for writing and submitting the manuscript.
%%WWM came up with the initial research concept and edited the manuscript.
%%OTS obtained the funding and edited the manuscript.
%%EBF provided the formal analysis and validation. He also edited the manuscript.
%%GEH Supervised the undergraduates, wrote the software and administers the project github and Zenodo repositories.
%%
%% Authors can use the Contributor Role Taxonomy (CRediT) at
%% for ideas on how write a good statement tailored to their needs.

\end{contribution}


\software{\textsc{numpy} \citep{numpy2020}, \textsc{matplotlib} \citep{mpl2007}, \textsc{seaborn} \citep{seaborn2021}, \textsc{h5py} \citep{h5py2021}, \textsc{colossus} \citep{colossus18} and \textsc{pyfftw} \citep{gomersall2021pyfftw}}


\section*{Data Availability} \label{section:acknowledgements}

The power spectra used in this analysis are provided alongside the online figures: \url{www.calvinosinga.com/papers/hi_cosmo/sup_analysis}. Data from specific figures in this publication are available through upon request via email to the authors. The IllustrisTNG data is available at \url{www.tng-project.org/data}.

\appendix

\section{Resolution Effects} \label{app:resolution}

\begin{figure*}
    \centering
    \includegraphics[width=\linewidth]{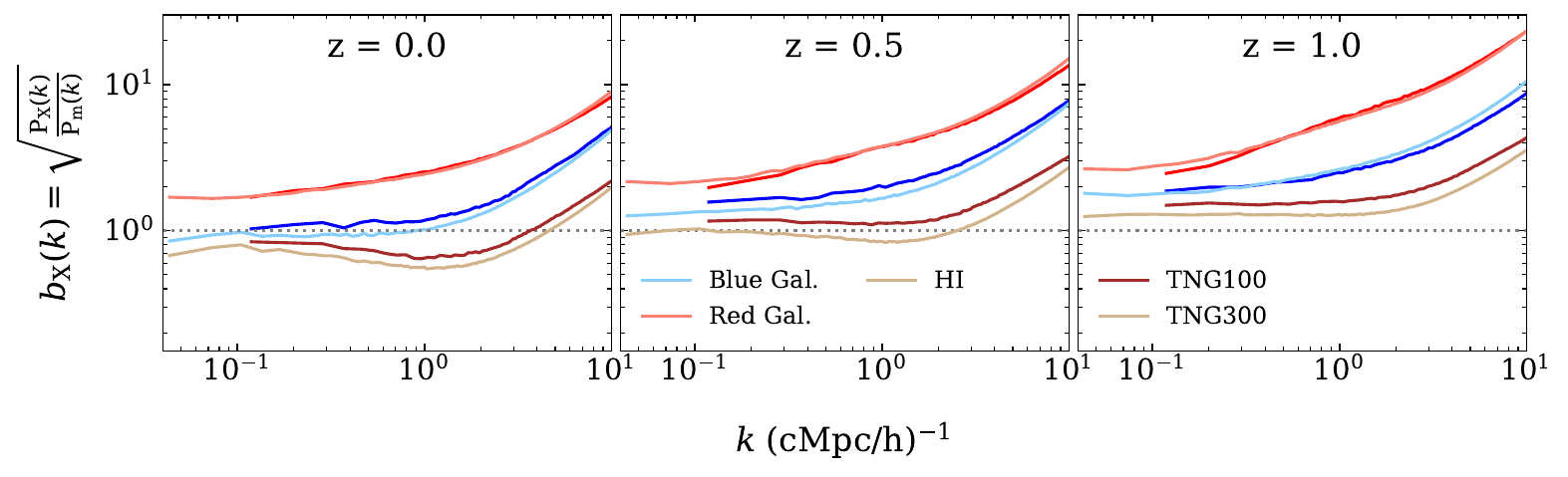}
    \caption{Comparisons of the \hi (brown) and galaxy (blue and red) biases from the primary volumes of TNG100 (dark colors) and TNG300 (light). The galaxy biases agree between the two resolutions, particularly at early redshifts and for red galaxies. Conversely, the TNG100 \hi bias is vertically offset by $\sim$25\% from TNG300, although their scale-dependencies agree. Our predictions of model errors are robust against the vertical offset because they are not dependent on the precise value of the \hi bias.}
    \label{fig:sim_comp}
\end{figure*}

\begin{figure*}
    \centering
    \includegraphics[width=\linewidth]{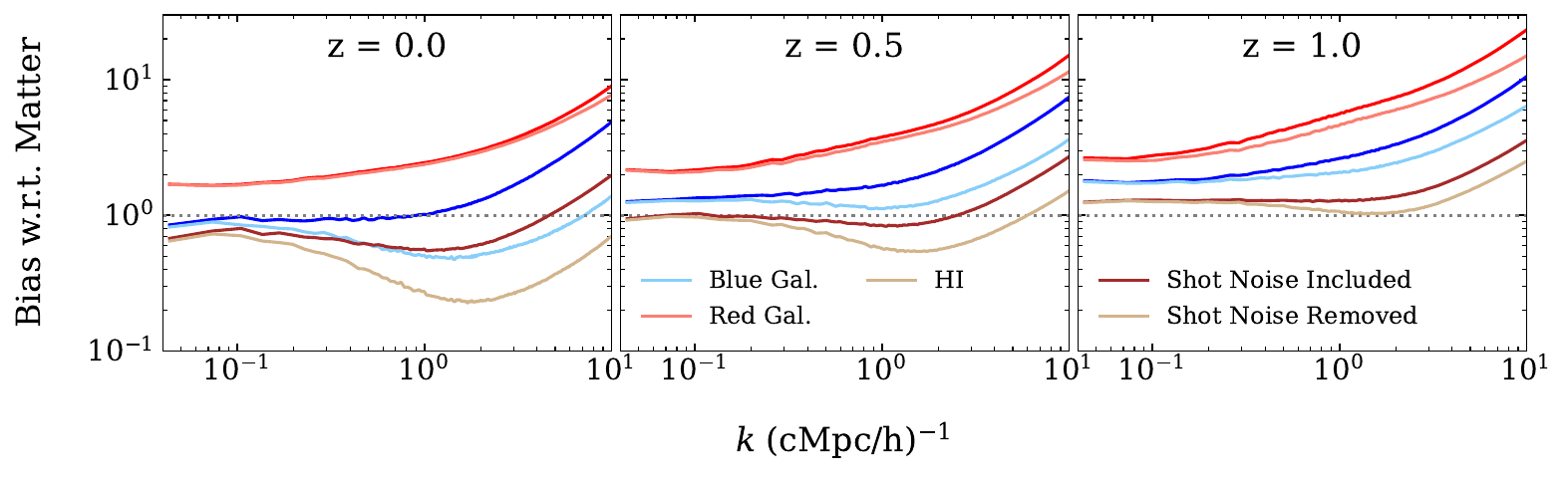}
    \caption{The contribution of shot noise to tracer biases, determined by comparing two bias definitions: our fiducial definition with shot noise included (darker colors, Equation~\ref{eq:bias}) and an alternative without shot noise (lighter colors, Equation~\ref{eq:bias_sn}). The large-scale values agree between the two definitions for all tracers at all redshifts, although the biases without shot noise are more scale-dependent. The differences between the definitions are small for large populations with small shot noise terms, such as red galaxies. Both definitions produce the same errors, although they can alter what errors we associate with assuming a constant bias. Definitions with shot noise mitigate the scale-dependence of the bias, which reduces constant-bias error.
    %The smaller samples of \hi and blue galaxies at later redshifts boost the shot noise, increasing the disparity between the definitions. The increased scale-dependence in the alternative definition without shot noise implies that our estimates of the error associated with assuming a constant bias (Section~\ref{ing:bias}) are mitigated by shot noise. These differences do not translate to the error estimates of the whole models (Section~\ref{sec:perf}) since models with either definition will include shot noise (see text).
    }
    \label{fig:shot_noise}
\end{figure*}

We study the resolution dependence of our results in Fig.~\ref{fig:sim_comp} by comparing the \hi, blue, and red galaxy biases from TNG100 and TNG300, which have resolutions that differ by approximately an order of magnitude \citep{nelson_illustristng_2019}. We find that red galaxy bias is nearly identical between the two resolutions within overlapping scales, and the blue galaxy bias converges for $z = 1$. At later redshifts, the blue galaxy bias has a slight vertical offset, similar to the \hi bias at all redshifts. The different behavior between the various tracers is expected (Section~\ref{disc:sims}) as the blue or \hi-riƒch galaxies tend to have masses closer to the resolution limit than red galaxies, particularly at late redshifts \citep[\citetalias{villaescusa-navarro_ingredients_2018},][]{nelson_first_2018}. Furthermore, star-formation rates are generally enhanced with better resolution, leading to less \hi overall \citep{dave_neutral_2013}.

Both the correlation coefficients and pairwise velocity dispersions (PVD) are also impacted significantly by resolution. The effects on the correlation coefficient can be inferred from Fig.~\ref{fig:sim_comp} --- the downward shift of the \hi bias implies a decrease in the \hi auto power spectrum, which causes all \hi-galaxy correlation coefficients to shift down as well (online figures). Similarly, all PVDs are expected to be dependent on resolution due to changing satellite fractions altering how virialized motions manifest in redshift space \citep{slosar_pairwise_2006}. Given these resolution effects, we cannot conclude that the bias, correlation coefficient, and PVD values presented in Section~\ref{sec:ing} are converged with resolution. 

However, our conclusions about model performance in Section~\ref{sec:perf} are independent of resolution effects. The precise values of the large-scale bias and correlation coefficient do not impact model error estimates, since they are removed in the $P_{\mathrm{model}} / P\shi$ ratios. Tracer RSDs can change at different mass resolutions due to how they sample the large-scale velocity field \citep{jennings_velocity_2015, robertson_modelling_2024}, but we find similar RSD errors in TNG100 \citepalias[][]{villaescusa-navarro_ingredients_2018}.
 
%%%%%%%%%%%%%%%%%%%%%%%%%%%%%%%%%%%%%%%%%%%%%%%%%%%%%%%%%%%%%%%%%%%%%%%%%%%%%%%%%%%%%%%%%%%%%%%%%%%%%%%%%%%%%

\section{Shot Noise} \label{app:SN}

Shot noise arises in power spectra due to the discrete sampling of a continuous density field. Generally, this term is thought to be negligible on large scales but can be significant on small scales \citep{castorina_spatial_2017}. We establish the role of shot noise in our results by first outlining its expected behavior and then comparing bias definitions with and without shot noise in Fig.~\ref{fig:shot_noise}.

Shot noise is inversely proportional to the sample size, like $P_{\mathrm{SN}} \sim 1 / n$ where $n$ is the sample number density. For this work, however, we use mass-weighted power spectra which has a different expression for the shot noise contribution:
\begin{equation}
    P_{\mathrm{SN}} = V \frac{\left\langle \sum\limits_i (M_i)^2 \right\rangle}{\left(\langle \sum\limits_i M_i \rangle \right)^2} \,.
\end{equation}
Here, $V$ is the volume and $M_i$ is the mass of the $i$th source, which in our case would be an \hi mass element or a galaxy. If the shot noise is purely Poisson, then $P_{\mathrm{SN}}$ is scale-independent, although $P_{\mathrm{SN}}$ can become slightly scale-dependent via residual correlations on the smallest scales \citep{chen_extracting_2021}.

Instead of folding the shot noise term into the bias like in Equation~\ref{eq:bias}, we can separate the two terms with an alternative bias definition \citep[see also equation 7 from][]{ando_redshift_2019}:
\begin{equation}
    \label{eq:bias_sn}
    b_i (k) = \frac{P_{i \times \mathrm{m}}(k)}{P_{\mathrm{m}} (k)} = \sqrt{\frac{P_i (k) - P_{\mathrm{SN}}}{P_{\mathrm{m}} (k)}} \,,
\end{equation}
for some tracer $i$. With this new bias definition, the relationship between the auto power spectra of matter and tracer $i$ becomes
\begin{equation}
    P_i (k) = b_i (k)^2 P_{\mathrm{m}} (k) + P_{\mathrm{SN}} \,.
    \label{eq:bias_sn_auto}
\end{equation}
We note that our fiducial models (Equations~\ref{eq:auto_full}-\ref{eq:cross_full}) and models like Equation~\ref{eq:bias_sn_auto} fundamentally include the same components, but attribute the shot noise to different terms. Consequently, our tests of the models using scale-dependent biases remain the same regardless of bias definition, although the errors we attribute to the bias term would change. We compare our fiducial bias definition to the bias from Equation~\ref{eq:bias_sn} in Fig.~\ref{fig:shot_noise} and re-interpret the errors linked to the bias in the context of this alternative bias definition.

All biases shown in Fig.~\ref{fig:shot_noise} receive negligible shot noise contributions on the largest scales, although shot noise plays a larger role in blue galaxies and \hi than red galaxies. We attribute this to the larger sample size and stronger clustering of red galaxies, both of which diminish the significance of shot noise such that $P_{\mathrm{Red}} > P_{\mathrm{SN}}$. At earlier redshifts, the red-galaxy shot noise is larger because there are fewer red galaxies. As more galaxies quench at later redshifts, the red population grows, suppressing shot noise. Conversely, the blue galaxy and \hi abundance decreases, strengthening shot noise.

If we were to exclude the shot noise from the bias using two distinct terms as in Equation~\ref{eq:bias_sn_auto}, we would attribute \textit{more} error to the bias term. The shot noise mitigates the scale-dependence of the bias such that assuming a constant bias appears better in Section~\ref{ing:bias} than it would with the alternative definition without the shot noise. This conclusion further illustrates that our results represent the best-case error estimates.

\section{Redshift Evolution} \label{app:addz}

\begin{figure}
    \centering
    \includegraphics[width=0.85\linewidth]{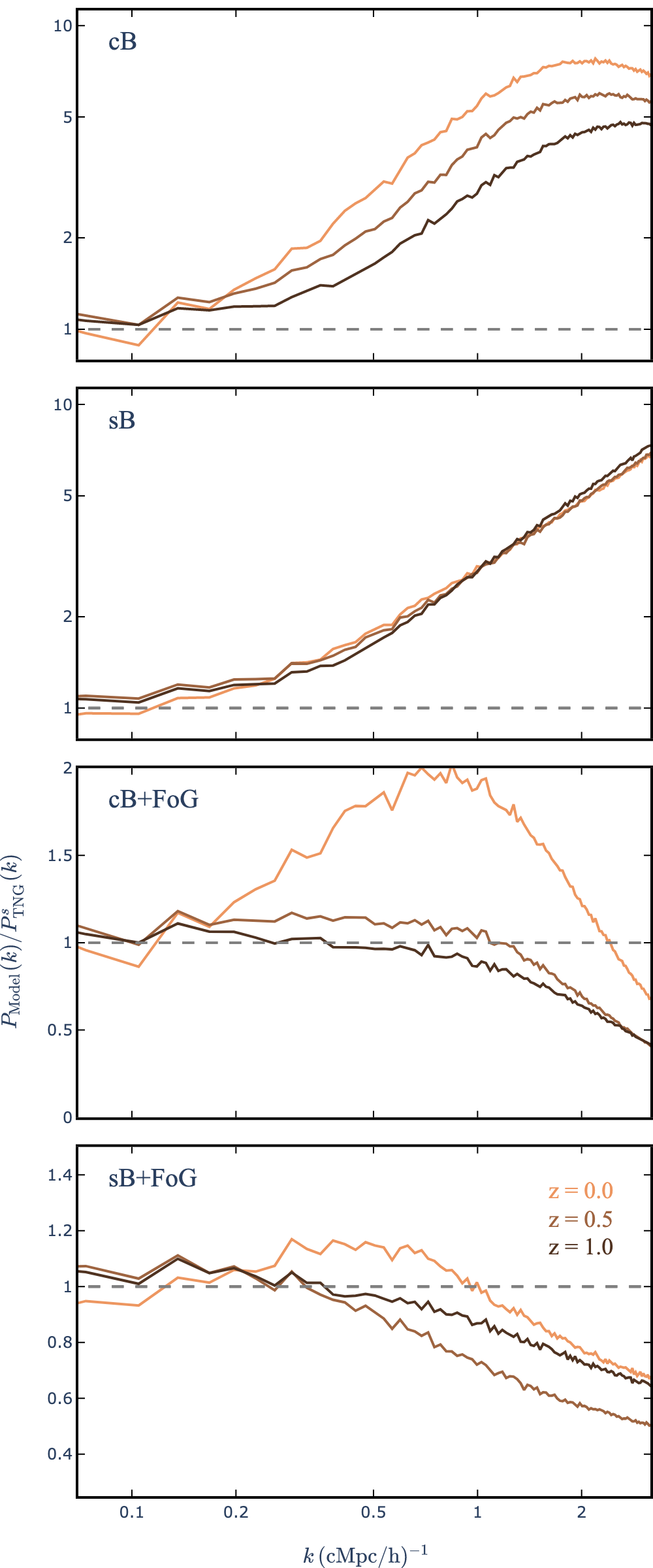}
    \caption{Redshift evolution of the ratio of the \hi auto power spectra from each model over the one calculated directly from TNG300, with earlier redshifts shown in darker colors. The top two panels have logarithmic y-axes due to their large errors. All models perform worse at later redshifts due to the increasing presence of nonlinearities and baryonic effects. The $z = 1$ \hi bias remains constant to small scales, allowing \xmC to agree within 10\% of the actual power spectrum until $k \sim 1$ \hmpc.}
    \label{fig:auto_err_all}
\end{figure}

\begin{figure}
    \centering
    \includegraphics[width=\linewidth, trim = {1cm 1cm 2.5cm 5cm}, clip]{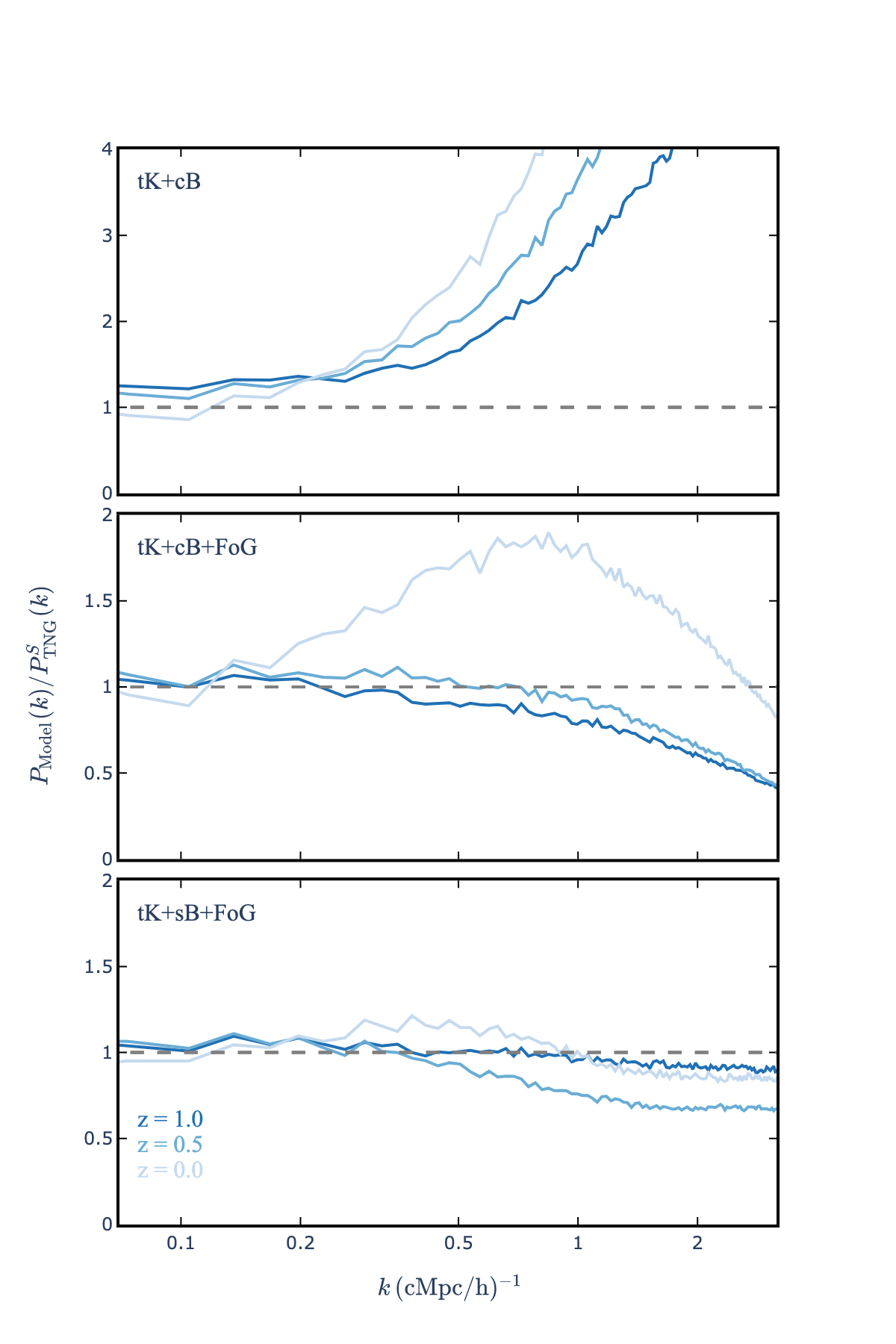}
    \caption{Redshift evolution of the ratio of \hib from each model over the one calculated directly from TNG300, with earlier redshifts shown in dark colors. All models perform worse at later redshifts, particularly at $z = 0$, similar to the trend in Fig.~\ref{fig:auto_err_all}.}
    \label{fig:xpk_err_all}
\end{figure}

In Section~\ref{sec:perf}, we examined the errors associated with various models for \hi auto and cross-power spectra at a single redshift. Here, we extend this analysis by characterizing some redshift trends in Fig.~\ref{fig:auto_err_all} and \ref{fig:xpk_err_all}. We relegate the complete set of \hi-galaxy cross-power spectra and $\Omega_{\mathrm{HI, fit}}/\Omega_{\mathrm{HI, TNG}}$ at all redshifts to the online figures as they do not yield significantly more insight into these trends.

All models in both the auto and cross power spectra perform worse at later redshifts due to stronger nonlinearities and baryonic effects. Both effects cause tracer biases to become more scale-dependent at larger scales and complicate how RSDs affect the clustering. This is particularly true at $z = 0$, when assuming a constant \hi bias introduces significant errors into the corresponding models (\amA, \amC, \xmB, and \xmC). The RSD errors in \amD and \xmD clearly manifest uniquely at $z = 0$ as compared to the other redshifts. Attributing this phenomenon to any one process is difficult, although we note that the \hi-halo mass relation and \hi density profiles change relatively rapidly between $z = 1$ and $z = 0$ \citepalias[see figures 4 and 5 from][]{villaescusa-navarro_ingredients_2018}. 

\section{Comparing Auto and Cross-Power Spectra} \label{app:ax_compare}

\begin{figure}
    \centering
    \includegraphics[width=\linewidth]{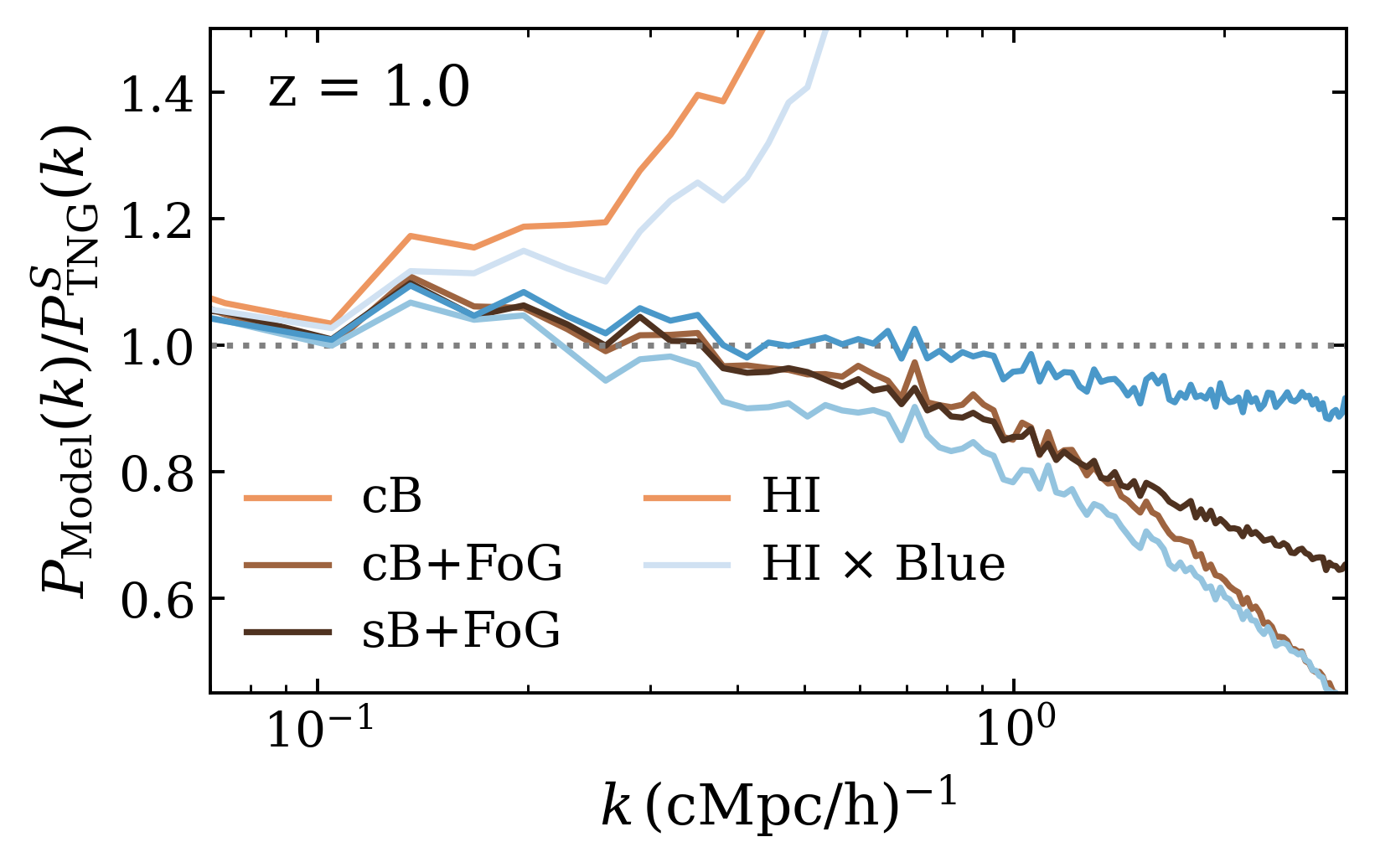}
    \caption{Comparison of the models for the \hi auto and \hib cross-power spectra at $z = 1$, with lighter colors corresponding to simpler models. \hib deviates from the cross-power spectrum comparably to or by less than equivalent \hi auto power spectrum models, suggesting that \hib may be easier to model than the \hi auto power spectra.}
    \label{fig:xvauto}
\end{figure}

In Section~\ref{sec:perf}, we studied the performance of auto and cross-power spectra separately. We can extend this analysis by comparing the best-performing cross-power spectra (\hib) to the \hi auto power spectra (Fig.~\ref{fig:xvauto}) to gain insight into which measurement might be analytically preferred. If \hib models outperform those of \hi auto power spectra, then \hib would yield more accurate constraints on cosmological \hi properties, assuming any additional systematic complications arising from cross-correlating different surveys are properly handled.

Fig.~\ref{fig:xvauto} demonstrates that the tested models for \hib outperform the equivalent \hi auto power spectrum models, or obtain comparable errors in the case of \amC and \xmC. The reason changes for each set of models. For example, \xmB improves on \amA because the FoG effect, which both models neglect, is weaker in blue galaxy clustering, reducing its overall contribution to \hib. The weaker FoG suppression in blue galaxies also mitigates the RSD errors in \xmD as compared to \amD.

The key takeaway from Fig.~\ref{fig:xvauto} is that cross-correlating \hi and galaxy distributions can mitigate the impact of complexities in \textit{both} auto power spectra, since any discrepancies between the model and the actual auto power spectra are squared but could cancel in the cross-power spectra. As an example of this, consider the $z \leq 0.5$ \hi and galaxy biases. Each exhibit opposite scale-dependencies such that the product of the \hi and galaxy bias appears more scale-independent than either individually, mitigating errors in the cross-power spectra relative to its auto-power counterparts. Fig.~\ref{fig:xvauto} motivates obtaining cosmological constraints from both \hi auto and \hi-galaxy cross-power spectra as a method of testing our analytical understanding of \hi distributions self-consistently. However, we note that this result arises from coincidental offsets that are not present at all redshifts (online figures) and leave further analysis on the impact of the offsets on constraints from auto and cross-power spectra to future work.

%% For this sample we use BibTeX plus aasjournalv7.bst to generate the

\bibliography{references, edit_references}{}
\bibliographystyle{aasjournalv7}



\end{document}